\documentclass[12pt,a4paper]{article}
\usepackage[dvips]{graphics}
\usepackage{cite}
\usepackage{amssymb,epsfig,pennames,rotating}
\newcommand{\inmath}[1] {\ifmmode#1\else$#1$\fi}
\newcommand{\definmath}[2] {\def#1{\ifmmode#2\else$#2$\fi}}
\definmath{\dEdx} {{\mathrm d}E/{\mathrm d}x}
\definmath{\dEdxn} {{\mathrm d} E/{\mathrm d}x _{\mathrm norm}}
\definmath{\ededx} {\sigma_{\dEdx}}
\newcommand{\alphas} {\alpha_{\mathrm{s}}}
\newcommand{\alphaem} {\mbox{$\alpha_{\mathrm{em}}$}}

\newcommand{\spi}   {\pi_s}
\definmath{\PWpm} {\mathrm{W}^{\pm}}      
\definmath{\Plp} {\ell^{+}}        
\definmath{\Plm} {\ell^{-}}        
\definmath{\Plpm}   {\ell^{\pm}}         
\definmath{\Pgtp} {\tau^{+}}        
\definmath{\Pgtm} {\tau^{-}}        
\definmath{\Pgtpm}   {\tau^{\pm}}         
\definmath{\Pgn}  {\nu}          
\definmath{\Pagn} {\overline{\nu}}     
\definmath{\Pf}      {\mathrm{f}}
\definmath{\Paf}  {\overline{\mathrm{f}}}
\definmath{\Pq}      {\mathrm{q}}
\definmath{\Paq}  {\overline{\mathrm{q}}}
\definmath{\Pu}      {\mathrm{u}}
\definmath{\Pau}  {\overline{\mathrm{u}}}
\definmath{\Pd}      {\mathrm{d}}
\definmath{\Pad}  {\overline{\mathrm{d}}}
\definmath{\Ps}      {\mathrm{s}}
\definmath{\Pas}  {\overline{\mathrm{s}}}
\definmath{\Pc}      {\mathrm{c}}
\definmath{\Pac}  {\overline{\mathrm{c}}}
\definmath{\Pb}      {\mathrm{b}}
\definmath{\Pab}  {\overline{\mathrm{b}}}
\definmath{\Pt}      {\mathrm{t}}
\definmath{\Pat}  {\overline{\mathrm{t}}}
\definmath{\Pap}  {\overline{\mathrm{p}}}
\definmath{\Pan}  {\overline{\mathrm{n}}}
\definmath{\PaD}  {\overline{\mathrm{D}}}
\definmath{\PaDz} {\overline{\mathrm{D}}^{0}}
\definmath{\PaB}  {\overline{\mathrm{B}}}
\definmath{\PaBz} {\overline{\mathrm{B}}^{0}}
\definmath{\PsDpm}   {\mathrm{D}^{\pm}_{\mathrm{s}}}  
\definmath{\PcgLpm}  {\Lambda^{\pm}_{\mathrm{c}}}  
\definmath{\PD} {\mathrm{D}}     
\definmath{\PDst} {\mathrm{D}^{*}}     
\definmath{\PDstp} {\mathrm{D}^{*+}}     
\definmath{\PDstm} {\mathrm{D}^{*-}}     
\definmath{\PgLz} {\Lambda^{0}}        

\newcommand{\nt}{\tilde{\chi}^0}
\newcommand{\massof}[1] {m_{\smash{#1}\mathstrut}}
\newcommand{\mtop}   {\massof{\mathrm{top}}}
\newcommand{\mHiggs} {\massof{\mathrm{Higgs}}}

\newcommand{\mPZ} {\massof{\mathrm{Z}}}

\newcommand{\Rb}  {R_{\mathrm{b}}}
\newcommand{\Rc}  {R_{\mathrm{c}}}

\newcommand{\AFBb}   {A_{\mathrm{FB}}^{\mathrm{b}}}
\newcommand{\AFBc}   {A_{\mathrm{FB}}^{\mathrm{c}}}
\newcommand{\AFBbSM}   {A_{\mathrm{FB}}^{\mathrm{b,\,SM}}}
\newcommand{\AFBcSM}   {A_{\mathrm{FB}}^{\mathrm{c,\,SM}}}
\newcommand{\AFB}    {A_{\mathrm{FB}}}

\newcommand{\Lsig}      {L/\sigma_{L}}
\newcommand{\epem}   {\rm e^+ e^-}

\newcommand{\PB} {\rm B}

\newcommand{\btol}      {\Pb\to\ell}

\newcommand{\btoctol}      {{\rm b}\to{\rm c}\to\ell}
\newcommand{\btocbartol}   {{\rm b}\to\overline{\rm c}\to\ell}
\newcommand{\btoccbartol}   {{\rm b}\to[{\rm c},\overline{\rm c}]\to\ell}
\newcommand{\btoccbartosp}   {{\rm b}\to[{\rm c},\overline{\rm c}]\to\spi}

\newcommand{\ctol}      {{\rm c}\to\ell}
\newcommand{\ctosp}      {{\rm c}\to\spi}

\newcommand{\btoctolsp}      {{\rm b}\to{\rm c}\to\ell/\spi}
\newcommand{\btocbartolsp}   {{\rm b}\to\overline{\rm c}\to\ell/\spi}
\newcommand{\btoccbartolsp}   {{\rm b}\to[{\rm c},\overline{\rm c}]\to\ell/\spi}
\newcommand{\ctolsp}      {{\rm c}\to\ell/\spi}

\newcommand{\btolshort}      {\Pb\hspace{-1.5pt}\to\hspace{-1.5pt}\ell}
\newcommand{\btoccbartolspshort}   {{\rm b}\hspace{-1.5pt}\to\hspace{-1.5pt}[{\rm c},\overline{\rm c}]\hspace{-1.5pt}\to\hspace{-1.5pt}\ell/\spi}
\newcommand{\ctolspshort}      {{\rm c}\hspace{-1.5pt}\to\hspace{-1.5pt}\ell/\spi}
\newcommand{\Rmis} {R_{\rm mis}}
\newcommand{\pt} {p_t}
\newcommand{\ptsq} {\pt^2}
\newcommand{\Psig} {{\cal P}_{\rm sig}}
\newcommand{\Psigvtx} {\Psig^{\rm (vtx)}}
\newcommand{\Psigl} {\Psig^{(\ell)}}
\newcommand{\Psigsp} {\Psig^{(\spi)}}
\newcommand{\Psiglsp} {\Psig^{(\ell/\spi)}}

\newcommand{\Psigbl} {\Psig^{(\Pb\to\ell)}}
\newcommand{\Psigbcl} {\Psig^{(\Pb\to[\Pc,\Pac]\to\ell)}}
\newcommand{\Psigcl} {\Psig^{(\Pc\to\ell)}}
\newcommand{\Psigbsp} {\Psig^{(\Pb\to\spi)}}
\newcommand{\Psigbcsp} {\Psig^{(\Pb\to[\Pc,\Pac]\to\spi)}}
\newcommand{\Psigcsp} {\Psig^{(\Pc\to\spi)}}

\newcommand{\AFBobs} {A_{\rm FB}^{obs}}
\newcommand{\epsb}   {(\epsilon_{\rm \Pb}-\overline{\epsilon}_{\rm \Pb})}
\newcommand{\epsc}   {(\epsilon_{\rm \Pc}-\overline{\epsilon}_{\rm \Pc})}
\newcommand{\epsuds} {(\epsilon_{\rm \Pu\Pd\Ps}-
                      \overline{\epsilon}_{\rm \Pu\Pd\Ps})}
\definmath{\GeV}  {\mathrm{GeV}}
\definmath{\GeVc} {\mathrm{GeV}\!/c}
\definmath{\GeVcc}   {\mathrm{GeV}\!/c^2}
\definmath{\MeV}  {\mathrm{MeV}}
\definmath{\MeVc} {\mathrm{MeV}\!/c}
\definmath{\MeVcc}   {\mathrm{MeV}\!/c^2}
\definmath{\MVm}  {\mathrm{MV}\!/\mathrm{m}}
\definmath{\keV}  {\mathrm{keV}}
\definmath{\keVcm}   {\mathrm{keV}\!/\mathrm{cm}}
\definmath{\kV}      {\mathrm{kV}}
\definmath{\km}      {\mathrm{km}}
\definmath{\meter}   {\mathrm{m}}
\definmath{\cm}      {\mathrm{cm}}
\definmath{\mm}      {\mathrm{mm}}
\definmath{\micron}  {\mu\mathrm{m}}
\definmath{\nm}      {\mathrm{nm}}
\definmath{\kg}      {\mathrm{kg}}
\definmath{\gram} {\mathrm{g}}
\definmath{\second}  {\mathrm{s}}
\definmath{\microsec}   {\mu\mathrm{s}}
\definmath{\degree}  {^\circ}
\definmath{\degC} {^\circ\mathrm{C}}
\definmath{\ohm}  {\Omega}
\definmath{\Mohm} {\mathrm{M}\Omega}
\definmath{\rad}  {\mathrm{rad}}
\definmath{\mrad} {\mathrm{mrad}}
\definmath{\nb}      {\mathrm{nb}}
\newcommand{\eqref}[1]  {(\ref{#1})}
\newcommand{\PhysLett}  {Phys.~Lett.}

\newcommand{\PhysRev}   {Phys.~Rev.}
\newcommand{\NPhys}  {Nucl.~Phys.}
\newcommand{\NIM} {Nucl.~Instrum.\ Methods}
\newcommand{\ZPhys}  {Z.~Phys.}
\newcommand{\EurPhysJ} {Eur.~Phys.~J.}

\newcommand{\OPALColl}    {OPAL Collaboration}
      \newcommand{\bb}      {{\rm      b\overline{\rm  b}}}

\definmath{\BR}{\mathrm B}
\definmath{\N2D}{N_{\PDst,\Pgpm}}

\definmath{\netel}{{\cal N}_{\rm el}}
\definmath{\netcv}{{\cal N}_{\rm cv}}
\definmath{\netmu}{{\cal N}_{\mu}}
\definmath{\netb}{{\cal N}_{\rm b}}
\definmath{\netc}{{\cal N}_{\rm c}}
\definmath{\netbc}{{\cal N}_{\rm bc}}
 
\definmath{\nf}{n_F}
\definmath{\nb}{n_B}
\definmath{\nt}{n_T}
\definmath{\mix}{\overline{\chi}}

  \def\rotninety{\special{ps:currentpoint currentpoint translate 90 rotate neg exch
                             neg exch translate}}

\definmath{\GeV}  {\mathrm{GeV}}
\definmath{\GeVc} {\mathrm{GeV}\!/c}
\definmath{\GeVcc}   {\mathrm{GeV}\!/c^2}
\definmath{\MeV}  {\mathrm{MeV}}
\definmath{\MeVc} {\mathrm{MeV}\!/c}
\definmath{\MeVcc}   {\mathrm{MeV}\!/c^2}
\definmath{\MVm}  {\mathrm{MV}\!/\mathrm{m}}
\definmath{\keV}  {\mathrm{keV}}
\definmath{\keVcm}   {\mathrm{keV}\!/\mathrm{cm}}
\definmath{\kV}      {\mathrm{kV}}
\definmath{\km}      {\mathrm{km}}
\definmath{\meter}   {\mathrm{m}}
\definmath{\cm}      {\mathrm{cm}}
\definmath{\mm}      {\mathrm{mm}}
\definmath{\micron}  {\mu\mathrm{m}}
\definmath{\nm}      {\mathrm{nm}}
\definmath{\kg}      {\mathrm{kg}}
\definmath{\gram} {\mathrm{g}}
\definmath{\second}  {\mathrm{s}}
\definmath{\microsec}   {\mu\mathrm{s}}
\definmath{\degree}  {^\circ}
\definmath{\degC} {^\circ\mathrm{C}}
\definmath{\ohm}  {\Omega}
\definmath{\Mohm} {\mathrm{M}\Omega}
\definmath{\rad}  {\mathrm{rad}}
\definmath{\mrad} {\mathrm{mrad}}
\definmath{\nb}      {\mathrm{nb}}
\definmath{\dEdx} {{\mathrm d}E/{\mathrm d}x}
\definmath{\dEdxn} {{\mathrm d} E/{\mathrm d}x _{\mathrm norm}}
\definmath{\ededx} {\sigma_{\dEdx}}
\def\cc{\ifmmode {{\mathrm c\bar{\mathrm c}}}
    \else {${\mathrm c\bar{\mathrm c}}$} \fi}
\def\bb{\ifmmode {{\mathrm b\bar{\mathrm b}}}
    \else {${\mathrm b\bar{\mathrm b}}$} \fi}
\def\qq{\ifmmode {{\mathrm q\bar{\mathrm q}}}
    \else {${\mathrm q\bar{\mathrm q}}$} \fi}
\definmath{\PWpm} {\mathrm{W}^{\pm}}      
\definmath{\Pgtp} {\tau^{+}}        
\definmath{\Pgtm} {\tau^{-}}        
\definmath{\Pgtpm}   {\tau^{\pm}}         
\definmath{\Pgn}  {\nu}          
\definmath{\Pagn} {\overline{\nu}}     
\definmath{\Pq}      {\mathrm{q}}
\definmath{\Paq}  {\overline{\mathrm{q}}}
\definmath{\Pf}      {\mathrm{f}}
\definmath{\Paf}  {\overline{\mathrm{f}}}
\definmath{\Pu}      {\mathrm{u}}
\definmath{\Pau}  {\overline{\mathrm{u}}}
\definmath{\Pd}      {\mathrm{d}}
\definmath{\Pad}  {\overline{\mathrm{d}}}
\definmath{\Ps}      {\mathrm{s}}
\definmath{\Pas}  {\overline{\mathrm{s}}}
\definmath{\Pc}      {\mathrm{c}}
\definmath{\Pac}  {\overline{\mathrm{c}}}
\definmath{\Pb}      {\mathrm{b}}
\definmath{\Pab}  {\overline{\mathrm{b}}}
\definmath{\Pt}      {\mathrm{t}}
\definmath{\Pat}  {\overline{\mathrm{t}}}
\definmath{\Pap}  {\overline{\mathrm{p}}}
\definmath{\Pan}  {\overline{\mathrm{n}}}
\definmath{\PaD}  {\overline{\mathrm{D}}}
\definmath{\PaDz} {\overline{\mathrm{D}}^{0}}
\definmath{\PaB}  {\overline{\mathrm{B}}}
\definmath{\PaBz} {\overline{\mathrm{B}}^{0}}
\definmath{\PsDpm}   {\mathrm{D}^{\pm}_{\mathrm{s}}}  
\definmath{\PcgLpm}  {\Lambda^{\pm}_{\mathrm{c}}}  
\definmath{\PcgL}  {\Lambda_{\mathrm{c}}}  
\definmath{\PsBz}  {\overline{\mathrm{B}^0_{\mathrm{s}}}} 
\definmath{\PbgL}  {\Lambda_{\mathrm{c}}}  
\definmath{\PD} {\rm D}     
\newcommand{\PDz} {\rm D^0}
\newcommand{\PDp} {\rm D^+}
\newcommand{\PDsp} {\rm D_s^+}
\definmath{\PDstpm} {{\mathrm{D}^{*\pm}}}     
\newcommand{\PZz}  {\rm Z^0}
\definmath{\PKshort}{{\mathrm{K}}_s^0}
\definmath{\PKm}{{\mathrm{K}^-}}
\definmath{\Ppip}{{\pi^+}}

\definmath{\PgLz} {{\Lambda^{0}}}        
\def\D0{\ifmmode {{\mathrm D^0}} \else {${\mathrm D^0}$}\fi}
\def\Z0{\ifmmode {{\mathrm Z^0}} \else {${\mathrm Z^0}$}\fi}

\definmath{\Zgamma}{{\PZz}/\gamma^*}

\definmath{\nf}{n_F}
\definmath{\nb}{n_B}
\definmath{\nt}{n_T}
\definmath{\mix}{\overline{\chi}}

\definmath{\chipos}{\chi_{\rm pos}}

\definmath{\Rb}{R_{\rm b}}
\definmath{\Rc}{R_{\rm c}}
\definmath{\Rq}{R_{\rm q}}

\def\etal{et al.}

\begin{document}
%
%
\begin{titlepage}
\begin{center}{\large   EUROPEAN LABORATORY FOR PARTICLE PHYSICS
}\end{center}\bigskip
\begin{flushright}
       CERN-EP/99-170   \\ 29 November 1999
\end{flushright}
\bigskip\bigskip\bigskip\bigskip\bigskip
\begin{center}
    \huge\bf\boldmath Measurements of $\Rb$, $\AFBb$, and $\AFBc$ \\ 
in ${\rm e^+ e^-}$ Collisions at $130 - 189$~GeV \\
\end{center}\bigskip\bigskip
\begin{center}{\LARGE The OPAL Collaboration
}\end{center}\bigskip\bigskip
\bigskip\begin{center}{\large  Abstract}\end{center}
The cross-section ratio
$\Rb = \sigma( \epem \to \bb )/\sigma( \epem \to \qq )$ 
and the bottom and charm 
forward-backward asymmetries $\AFBb$ and $\AFBc$ 
are measured using event samples collected by the OPAL detector  
at centre-of-mass energies between 130 and 189 GeV.
Events with bottom quark production are selected with
a secondary vertex tag, and a hemisphere charge algorithm is used to 
extract $\AFBb$.
In addition, the bottom and charm 
asymmetries are measured using leptons from semileptonic decays of heavy
hadrons and pions from $\PDstp\to\PDz\Ppip$ decays.
The results are in agreement with the Standard Model predictions.
\bigskip\bigskip\bigskip\bigskip
\bigskip\bigskip
\begin{center}{\large
(submitted to Eur.~Phys.~J.~C)
}\end{center}
\end{titlepage}
\begin{center}{\Large        The OPAL Collaboration
}\end{center}\bigskip
\begin{center}{
G.\thinspace Abbiendi$^{  2}$,
K.\thinspace Ackerstaff$^{  8}$,
P.F.\thinspace Akesson$^{  3}$,
G.\thinspace Alexander$^{ 23}$,
J.\thinspace Allison$^{ 16}$,
K.J.\thinspace Anderson$^{  9}$,
S.\thinspace Arcelli$^{ 17}$,
S.\thinspace Asai$^{ 24}$,
S.F.\thinspace Ashby$^{  1}$,
D.\thinspace Axen$^{ 29}$
% ...
}\end{center}\bigskip
\bigskip
$^{  1}$School of Physics and Astronomy, University of Birmingham,
Birmingham B15 2TT, UK
\newline
$^{  2}$Dipartimento di Fisica dell' Universit\`a di Bologna and INFN,
I-40126 Bologna, Italy
\newline
$^{  3}$Physikalisches Institut, Universit\"at Bonn,
D-53115 Bonn, Germany
\newline
% ...
$^{  8}$CERN, European Organisation for Particle Physics,
CH-1211 Geneva 23, Switzerland
\newline
$^{  9}$Enrico Fermi Institute and Department of Physics,
University of Chicago, Chicago IL 60637, USA
\newline
% ...
$^{ 16}$Department of Physics, Schuster Laboratory, The University,
Manchester M13 9PL, UK
\newline
$^{ 17}$Department of Physics, University of Maryland,
College Park, MD 20742, USA
\newline
% ...
$^{ 23}$Department of Physics and Astronomy, Tel Aviv University,
Tel Aviv 69978, Israel
\newline
$^{ 24}$International Centre for Elementary Particle Physics and
Department of Physics, University of Tokyo, Tokyo 113-0033, and
Kobe University, Kobe 657-8501, Japan
\newline
% ...
$^{ 29}$University of British Columbia, Department of Physics,
Vancouver BC V6T 1Z1, Canada
\newline
% ...

\clearpage
\section{Introduction}
\label{sec:intro}

Many measurements of heavy quark production have been performed 
in $\epem$ collisions on the $\PZz$ resonance~\cite{bib:ew}.  
Among the
measured parameters are the production cross-sections of bottom 
and charm quark 
pairs relative to the hadronic cross-section, $\Rb$ and $\Rc$, and the
forward-backward asymmetries $\AFBb$ and $\AFBc$.
In this paper, measurements of $\Rb$, $\AFBb$, and $\AFBc$ at energies
above the $\PZz$ resonance are presented for 
$\epem \to \Zgamma \to \qq$ events, where
the effective centre-of-mass energy $\sqrt{s'}$ 
after initial-state radiation is required to satisfy 
$\sqrt{s'/s}>0.85$.
Similar measurements have been performed previously by other 
collaborations~\cite{bib-ALEPH_LEP2,bib-DELPHI_LEP2}.

The data collected with the OPAL detector at
LEP at centre-of-mass energies between 
$130$~GeV and $189$~GeV are analysed.
The basic techniques  are 
similar to those adopted in previous OPAL
measurements~\cite{bib:rb_172,bib:rb_183,bib-OPALlafb}.
The $\Rb$ measurement is based on the selection of a 
sample enriched in $\rm b \overline{b}$ events 
obtained with a secondary vertex tagging technique.
Both $r\phi$ and $rz$ information\footnote{The OPAL coordinate 
system is defined as a right-handed Cartesian coordinate system, with the 
$x$ axis pointing in the plane of the LEP collider towards the 
centre of the ring, the $z$ axis in the direction of the 
outgoing electrons, and $\theta$ and $\phi$ defined as the usual spherical 
polar coordinates.} from the silicon microvertex detector
are used in the selection of $\bb$ events, improving 
on~\cite{bib:rb_172,bib:rb_183} where only
$r\phi$ information was used.
This provides higher efficiency at comparable purity, thus enhancing the
statistical precision of the measurement.

Two measurements of forward-backward asymmetries are performed.
In the events selected by secondary vertex tagging, the 
asymmetry $\AFBb$ is measured using 
a hemisphere charge technique to identify the direction of emission 
of the primary quark.  An independent measurement of both the 
bottom and charm forward-backward asymmetries is performed using  
leptons from semileptonic decays of heavy hadrons and pions from 
$\PDstp\to\PDz\Ppip$ decays.  The asymmetries  obtained with 
the hemisphere charge method and the
lepton and slow pion methods are combined.

In Section~\ref{sec:evsel}, a brief
description of the OPAL detector and the event selection is given. 
The $\Rb$ measurement is discussed in Section~\ref{rb.sec}, followed by
a description of the asymmetry measurements in Section~\ref{afb.sec}.
Systematic errors on both the $\Rb$ and asymmetry measurements are given
in Section~\ref{systunc.sec}, and in Section~\ref{conclusions.sec},
all results are summarised.


\section{The OPAL Detector and Event Selection}
\label{sec:evsel}

A detailed description of the OPAL detector can be found 
elsewhere~\cite{bib-OPALdetectoratLEPtwo}. For this analysis, 
the most relevant parts of the 
detector are the silicon microvertex detector, the
tracking chambers, the
electromagnetic and hadronic calorimeters, and the muon chambers. 
The microvertex detector is essential for the reconstruction of secondary
vertices. The central 
detector provides precise measurements of the momenta of charged particles
by the curvature of their trajectories 
in a magnetic field of $0.435\;$T. In addition, it allows an identification
of charged particles through a combination of the measurement of the
specific energy loss $\dEdx$ and the momentum.
The electromagnetic calorimeter consists
of approximately 12000 lead glass blocks, which completely cover 
the azimuthal range up to polar angles
of $|\cos \theta|<0.98$. Nearly
the entire detector is surrounded with four layers of 
muon chambers, after approximately one metre of iron
from the magnet return yoke, which is instrumented as a hadron calorimeter.

Starting in 1995, the LEP experiments 
have collected data at increasing energies well above the $\PZz$ peak.
In this paper, the energy points are classified in five different sets, at 
centre-of-mass energies which will be generically called 
$\sqrt{s}$=133, 161, 172, 183, and 189~GeV.
Table~\ref{tab:lep_lumi_MC} shows the luminosity-weighted 
mean centre-of-mass energies
at which data were taken, and the corresponding integrated luminosities.
Additionally, calibration data taken at the $\PZz$ peak during 1996, 1997,
and 1998 are used to cross-check the analyses.

\begin{table}[htb]
\begin{center}
\begin{tabular}{|l|c|c|c|c|c|}
\hline
Dataset & 133 GeV & 161 GeV & 172 GeV & 183 GeV & 189 GeV \\ \hline \hline
$\langle\sqrt{s}\rangle$ (GeV)       
        & 133.3 & 161.3 & 172.1 & 182.7 & 188.6 \\
Integrated luminosity (pb$^{-1}$)    
        & 9.9 &  9.6 & 9.9  & 55.8 &  177 \\
\hline
\end{tabular}
\caption{The effective centre-of-mass energies and integrated luminosities
of data collected above the $\PZz$ peak.  The uncertainties on these
quantities have a negligible effect on the analyses presented in this paper.}
\label{tab:lep_lumi_MC}
\end{center}
\end{table}
 
Hadronic events, $\rm e^+e^- \rightarrow q \overline{q}$,  
are selected based on the 
number of reconstructed charged 
tracks and the energy deposited in the calorimeters. 
The selection of the subsample of non-radiative hadronic events, defined 
by the requirement $\sqrt{s'/s} > 0.85$, and the identification and
rejection of $\rm W^+W^-$ background are 
described in detail in~\cite{bib:rb_183}.
The remaining contamination from radiative hadronic events with true effective
centre-of-mass energy below $0.85\sqrt{s}$ is $5-10 \%$,  
depending on the centre-of-mass energy.
The residual contamination from four-fermion events (mainly W and Z pairs)
is largest at $\sqrt{s}=189\ \GeV$, where it is about 8$\%$.
These backgrounds to non-radiative hadronic $\qq$ events are accounted
for in the measurements.

Jets are reconstructed using the JADE algorithm
\cite{bib:jade} with the E0 recombination scheme~\cite{bib:E0},
keeping the invariant mass cut-off 
$x_{\rm min}=49\ \GeV^2$ fixed at all 
centre-of-mass energies.

Hadronic events are simulated using the PYTHIA Monte Carlo 
generator~\cite{bib:pythia}.
Heavy quark fragmentation is modelled according to 
the scheme by Peterson~\etal~\cite{bib:peterson} with fragmentation parameters 
tuned according to
the results in~\cite{bib:lep_heavy}.
Four-fermion background events are simulated with the grc4f 
generator~\cite{bib:grc4f}.
The events are passed through a detailed simulation of the OPAL 
detector~\cite{bib-GOPAL} before being analysed 
using the same procedure as for the data.

\section{Measurement of \boldmath$\Rb$\unboldmath}
\label{rb.sec}

The tagging of
$\bb$ events is based on the long lifetime ($\sim$1.5 ps) 
and hard fragmentation
of b-flavoured hadrons, which give rise to secondary vertices 
significantly displaced from the primary vertex.
The secondary vertex tag described in Section~\ref{sec:vtx} 
allows a clean and efficient reconstruction of $\bb$ events.
In Section~\ref{sec:rb}, the measurement of $\Rb$ with vertex
tagged events is described.

\subsection{\boldmath Secondary Vertex Tag} 
\label{sec:vtx}

The algorithm used for secondary vertex reconstruction is 
described in~\cite{bib:3d_btsvtx}.
For the results presented here, a three-dimensional vertex tagging algorithm 
is used, which takes advantage of the precise $z$ information 
provided by the OPAL microvertex detector.

The primary vertex
in each event is reconstructed as described in~\cite{bib:rb_new},
incorporating the average beam spot position determined 
from the measured tracks and the LEP beam-orbit measurements as a constraint. 
Although the beam spot is less precisely determined 
at energies above the $\PZz$ resonance than at the $\PZz$ peak, the 
resulting error on the
primary vertex position is still small compared to the error on the
reconstructed secondary vertex position.
The angular
acceptance is restricted to $|\cos\theta_T|<0.9$, where $\theta_T$ denotes
the polar angle of the thrust axis of the event.
Charged tracks used to reconstruct secondary vertices are selected 
as described in~\cite{bib:rb_172}
and least three of them are required to form a vertex.

Each hadronic event is divided into two 
hemispheres by the plane perpendicular to the thrust axis
and containing the nominal interaction point.
For each reconstructed secondary vertex the signed decay length
$L$ is defined as the distance
between the secondary and the primary vertex.
$L$ is taken to be positive if the secondary vertex is 
in the hemisphere pointed at  
by the momentum vector of the jet
which contains the vertex, and negative otherwise.
The decay length significance $L/\sigma_L$ is defined as the 
ratio of the decay length
and its error. 
Secondary vertices with $L/\sigma_L > 8$ are used to tag $\bb$ events.
This cut represents 
the best compromise between tagging efficiency and 
purity.

\subsection{\boldmath Determination of $\Rb$ from  Vertex-Tagged Events} 
\label{sec:rb}

\begin{figure}[p]
\begin{center}
\hspace{272pt}{\huge\bf OPAL}
\end{center}
\vspace{-122pt}
\begin{center}
\epsfig{file=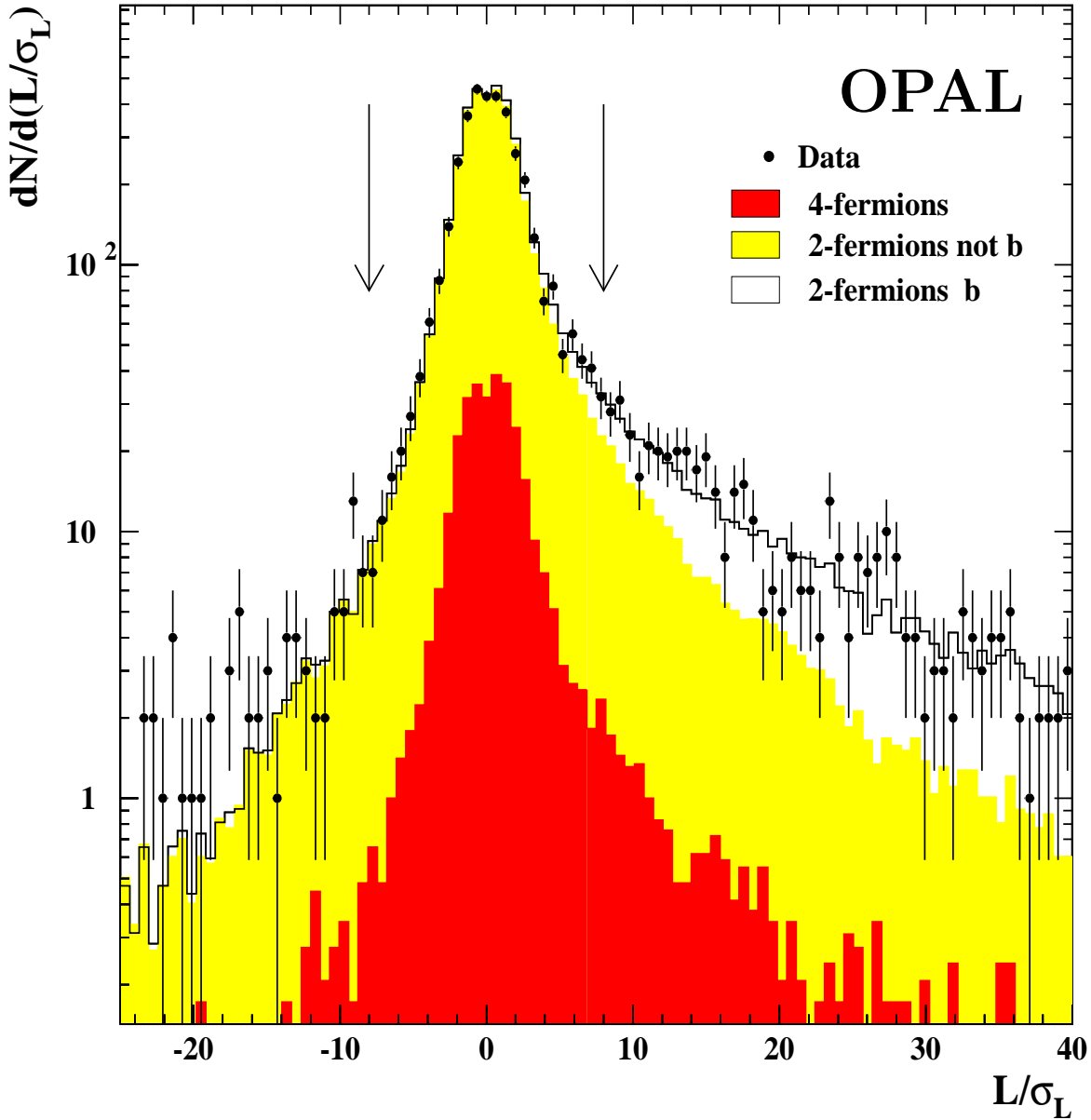,
width=1.0\textwidth,height=18.2cm} 
\end{center}
\vspace{-0.5cm}
\caption
{The decay length significance distribution for selected events
at 189 GeV centre-of-mass energy.
Only the most significant value is used for each event.
The points with error bars represent the data.
The dark-shaded histogram represents the expected contribution from 
four-fermion background, and the light-shaded area corresponds to the 
background from non-$\bb$ two-fermion events.  The unshaded histogram
indicates the b$\rm \overline{b}$-content in the sample.
The arrows show the position of the cut.} 
\label{fig:lsig_189}
\end{figure}

For the measurement of $\Rb$, the number of events tagged by a 
secondary vertex is determined and corrected for tagging 
efficiency and background.
Due to the limited statistics compared with the data collected at the 
$\PZz$ peak, a double tag technique as e.g.~in~\cite{bib-Rb} cannot be applied.
In order to reduce the sensitivity of the analysis to the detector 
resolution, a folded tag technique~\cite{bib:rb_new} 
is used: a hemisphere is assigned a tag 
if it contains a secondary
vertex with a decay length significance $\Lsig > 8$, or 
an anti-tag\footnote{In the literature, tagged and anti-tagged events
are sometimes referred to as ``forward'' and ``backward'' tagged events.
This convention is not used here to avoid confusion with the distinction
between forward and backward event hemispheres in the asymmetry analyses.}
if it contains a vertex with a decay length significance 
$\Lsig < -8$.
The number of anti-tagged hemispheres is then subtracted from 
the number of tagged
hemispheres.
After subtraction of the four-fermion background,
the difference between
the number of tagged and anti-tagged
hemispheres, $N - \overline{N}$, in a sample of $N_{\mathrm{had}}$ 
hadronic events can be expressed as
\begin{equation} \nonumber
{N-\overline{N} = 
      2N_{\mathrm{had}}[\epsb\Rb + \epsc\Rc +\epsuds(1-\Rb - \Rc)] \ , }
\nonumber
\end{equation}
where $\epsb$, $\epsc$, and $\epsuds$ are the differences between the
tagging and anti-tagging efficiencies for a given quark flavour.
The tagging efficiencies for $\rm u\overline{u}$, $\rm d\overline{d}$, and
$\rm s\overline{s}$ events are averaged (uds), since they are very similar.
For $\Rc$, 
the prediction of ZFITTER~\cite{bib:zfitter} is used. 
The four-fermion background is determined from  Standard Model production
cross-sections and from selection efficiencies estimated from 
Monte Carlo simulation.
The b-purity for the sample with the folded tag is 
defined as the fraction of $\rm b\overline{b}$
event hemispheres contained in the sample $N-\overline{N}$, and is about 
75$\%$.
The distribution of the decay length significance 
$L/\sigma_{L}$ is shown in Figure~\ref{fig:lsig_189}
for events at 189 GeV centre-of-mass energy,
together with the expectation from the Monte Carlo simulation.
The relative difference between data and Monte Carlo is 
below 1$\%$ for the folded-tag rate and $(8\pm4)\%$ for the 
anti-tag rate, which is more sensitive to 
modelling of the detector resolution.  
For $L/\sigma_L <  -10$, the number of 
events predicted by Monte Carlo differs by two 
standard deviations from the number of events in the data.
This indicates an incomplete simulation of the detector resolution, 
which is taken into account in the systematic errors as
described in Section~\ref{detres.subsubsec} below.
The agreement between data and simulation
is similar at the other centre-of-mass energies.

The selected event sample contains a $5-10\%$ contamination of
radiative hadronic events, with one or more energetic photons 
emitted in the initial state. In these events the 
effective centre-of-mass energy is reduced to values below 
$\sqrt{s'/s} = 0.85$ where
the predicted value of $\Rb$ is up to 30$\%$ larger.
In addition, the selection of non-radiative events is around $3\%$
less efficient for $\rm b\overline{b}$ final states than for other
flavours, because of a larger missing energy due to 
neutrinos in semileptonic b hadron decays.
The results are corrected for these effects, which are
estimated from Monte Carlo simulation. 
The measured values of $\Rb$ 
are also corrected for interference 
between initial and final-state radiation as described 
in~\cite{bib:rb_172}. This correction comes from the fact that 
Monte Carlo, which is used to model the data, does not contain 
interference between initial and final-state radiation.


The numbers of selected events and of tagged and anti-tagged hemispheres 
are given in Table~\ref{tab:Rb}.
The differences in hemisphere tagging efficiency, also listed in 
Table~\ref{tab:Rb}, have been determined from Monte Carlo simulation. 
Their errors include all the systematic 
uncertainties that will be described in Section~\ref{sec:sys}.
No systematic uncertainties other than those due to Monte Carlo statistics 
and detector resolution are assigned to the efficiencies in 
$\rm u\overline{u}$, 
$\rm d\overline{d}$, and $\rm s\overline{s}$ events,
as they represent a small fraction of the tagged sample.
The systematic error is dominated by the uncertainties 
from the event selection, the modelling of b and c fragmentation and decay,
and from the simulation of the detector resolution. 

\begin{table}[htb]
\begin{center}
\begin{tabular}{|c||c|c|c||c||c||c|}
\hline
  Energy & Events & $N$ & $\overline{N}$ 
  & \multicolumn{1}{c||}{\begin{tabular}{@{}c@{}}
                           Tagging efficiency\\
                           differences
                         \end{tabular}}
  & $\Rb$ & $R_{\rm b}^{\rm SM}$ 
  \rule[-3ex]{0pt}{7.5ex}\\
\hline
\hline
  133 GeV 
 & 
  \enspace 745      
 & 
  153      
 &   
  \enspace 11     
 &
  $\begin{array}{@{}r@{\,=\,}l@{}}
    \epsilon_{\rm \Pb}' & 0.347\enspace\pm0.026 \\
    \epsilon_{\rm \Pc}' & 0.045\enspace\pm0.009 \\
    \epsilon_{\rm \Pu\Pd\Ps}' & 0.0035\pm0.0014 
   \end{array}$
 & 
  0.190$\pm$0.023$\pm$0.007  
 &
  0.184
 \\
\hline
  161 GeV 
 & 
  \enspace 347     
 & 
  \enspace 58 
 &       
  \enspace\enspace 4     
 &  
  $\begin{array}{@{}r@{\,=\,}l@{}}
    \epsilon_{\rm \Pb}' & 0.331\enspace\pm0.013 \\
    \epsilon_{\rm \Pc}' & 0.045\enspace\pm0.002 \\
    \epsilon_{\rm \Pu\Pd\Ps}' & 0.0090\pm0.0010 
   \end{array}$
 &
  0.195$\pm$0.035$\pm$0.007 
 &
  0.171  
 \\
\hline
  172 GeV 
 & 
  \enspace 228    
 & 
  \enspace 23    
 &
  \enspace\enspace 3     
 & 
  $\begin{array}{@{}r@{\,=\,}l@{}}
    \epsilon_{\rm \Pb}' & 0.338\enspace\pm0.012 \\
    \epsilon_{\rm \Pc}' & 0.041\enspace\pm0.002 \\
    \epsilon_{\rm \Pu\Pd\Ps}' & 0.0066\pm0.0010 
   \end{array}$
 & 
  0.091$\pm$0.034$\pm$0.005
 &
  0.168 
 \\
\hline
  183 GeV 
 & 
  1186   
 & 
  232    
 &
  \enspace 25     
 &
  $\begin{array}{@{}r@{\,=\,}l@{}}
    \epsilon_{\rm \Pb}' & 0.351\enspace\pm0.009 \\
    \epsilon_{\rm \Pc}' & 0.048\enspace\pm0.002 \\ 
    \epsilon_{\rm \Pu\Pd\Ps}' & 0.0084\pm0.0008
   \end{array}$
 & 
  0.213$\pm$0.020$\pm$0.009
 &
  0.165
 \\
\hline
  189 GeV 
 & 
  3209   
 & 
  551    
 &
  \enspace 87    
 &
  $\begin{array}{@{}r@{\,=\,}l@{}}
    \epsilon_{\rm \Pb}' & 0.363\enspace\pm0.008 \\
    \epsilon_{\rm \Pc}' & 0.053\enspace\pm0.002 \\ 
    \epsilon_{\rm \Pu\Pd\Ps}' & 0.0099\pm0.0008
   \end{array}$
 & 
  0.158$\pm$0.012$\pm$0.007
 &
  0.164 
 \\
\hline
\end{tabular}
\caption{\label{tab:Rb}
The numbers of selected non-radiative events, tagged ($N$) and 
anti-tagged ($\overline{N}$) hemispheres, and tagging efficiency
differences 
$\rm \epsilon_{\rm q}' = (\epsilon_{\rm q}-\overline{\epsilon}_{\rm q})$
in the $\Rb$ analysis. The errors on the tagging efficiencies include 
systematic uncertainties.
The $\Rb$ results with the statistical and 
systematic errors and their Standard Model expectations are given 
in columns 6 and 7.}
\end{center}
\end{table}   

The dependence of the result on the assumed value of $\Rc$ 
can be parametrised as 
\begin{equation}
\label{rbdependence.eqn}
  \Delta \Rb = b \; \left( \Rc - R_{\rm c}^{\rm SM} \right)
\,.
\end{equation}
The parameter $b$ has been determined separately for each 
centre-of-mass energy.  Its values are given in Table~\ref{rbdependence.table}.
\begin{table}[htb]
\begin{center}
$\begin{array}{|l||c|c|c|c|c|}
 \hline
{\rm Dataset}  &  
133\, \GeV &
161\, \GeV &
172\, \GeV &
183\, \GeV &
189\, \GeV \rule[-1.5ex]{0pt}{4.5ex}
\\
\hline \hline
b \equiv \left( \Delta \Rb \right) / \left( \Rc - R_{\rm c}^{\rm SM} \right) &
 -0.12 &
 -0.11 &
 -0.11 &
 -0.12 &
 -0.13 \rule[-1.5ex]{0pt}{4.5ex}
\\
\hline
R_{\rm c}^{\rm SM} &
0.223 &
0.244 &
0.249 &
0.253 &
0.255 \rule[-1.5ex]{0pt}{4.5ex}
\\
\hline
\end{array}$
\caption{\label{rbdependence.table}
The dependence of the measured $\Rb$ value on the assumed value of 
$\Rc$ for each centre-of-mass energy.  The parameter $b$ gives
the change in $\Rb$ for an assumed deviation of $\Rc$ from its Standard
Model prediction.}
\end{center}
\end{table}

As a cross-check, the analysis is repeated on calibration data collected
at the $\PZz$ peak. A value of
$\Rb(\sqrt{s}=m_{\PZz}) = 0.221 \pm 0.002({\rm stat.}) \pm 0.010({\rm syst.})$ 
is obtained.
The systematic error has been determined as for the high-energy data samples.
This result agrees within the errors with the LEP1
combined value of $\Rb^0 = 0.21664\pm0.00076$~\cite{bib:ew} 
and can be regarded as a
check of the evaluation of systematic errors for the measurements at
energies above the $\PZz$ peak.
Note that the $4 \%$ systematic error which is assigned to the
$\Rb$ measurement at $\sqrt{s} = 189$~GeV is larger than the $2 \%$ 
difference with respect to the LEP1 average which is observed at 
$\sqrt{s} = m_{\PZz}$.


\section{Measurement of Forward-Backward Asymmetries}
\label{afb.sec}

For the measurement of heavy quark 
forward-backward asymmetries, it is necessary to distinguish
the event hemispheres of the primary quark and antiquark
in addition to the quark flavour tagging.  Two complementary techniques are 
used.  
The first analysis provides a measurement of $\AFBb$
for the events that have been tagged by the presence
of a secondary vertex.  A hemisphere charge method is used 
to distinguish between quark and anti-quark hemispheres.

The second technique is used for a simultaneous measurement of 
$\AFBb$ and $\AFBc$.  It is based on the identification 
of leptons from semileptonic decays of heavy hadrons (``prompt leptons'')
and pions from $\PDstp\to\PDz\Ppip$ decays (``slow pions'').
The charge of these particles provides a clean distinction
between the primary quark and anti-quark hemispheres.

In Section~\ref{sec:afb}, the measurement of $\AFBb$ with
the hemisphere charge technique is described.
After a discussion of the lepton and slow pion identification in 
Section~\ref{leptid.subsec} and the flavour separation of the tagged 
samples (Section~\ref{flavoursep.subsec}), 
the asymmetry measurement based on prompt leptons
and slow pions is presented in Section~\ref{fit.subsec}.  
Finally, the combination of the two measurements is
treated in Section~\ref{combination.subsec}.

\subsection{\boldmath Measurement of $\AFBb$ with a 
Hemisphere Charge Technique} 
\label{sec:afb}

To obtain a sample enriched in $\rm b\overline{b}$ events suitable for the 
$\AFBb$ measurement, 
the secondary vertex algorithm used for the $\Rb$ analysis 
is employed. The analysis is limited to the range $|\cos\theta_T|<0.9$. 
Only tagged events are used, since Monte Carlo studies show that with the 
current statistics, the folded tag would result in a larger 
error on the forward-backward asymmetry. 
Thus, an event is considered if it contains a secondary 
vertex with a decay length significance $\Lsig > 8$ and no other
secondary vertex with $L/\sigma_L < -8$.
This cut is chosen to minimize the overall error. 

Each non-radiative hadronic event is divided into two hemispheres
by the plane perpendicular to the thrust axis that contains the nominal
interaction point.
Monte Carlo simulation shows that the direction of the thrust axis 
is a good approximation of 
the direction of emission of the initial $\rm q\overline{q}$ pair.
For each hemisphere, the hemisphere charge
$Q_{\rm hem}$ is computed as
\begin{equation}
\label{defjq}
 Q_{\rm hem}=\frac{\sum^{n}_{i=1}|p_{i}|^{\kappa}\cdot Q_{i}}
{\sum_{i=1}^{n} |p_{i}|^{\kappa}},
\nonumber
\end{equation}
where the sum runs over all $n$ tracks in the hemisphere, 
$p_{i}$ is the momentum component of track $i$ along the thrust axis,
$Q_{i}$ denotes its charge, and  $\kappa=0.4$ is a parameter tuned 
on simulated events 
for an optimal charge identification of a primary b or $\Pab$. 
To ensure a good hemisphere charge reconstruction, only events with 
more than three charged tracks per hemisphere are used.
Events are classified according to the sign of the difference $Q_F-Q_B$  
between the forward ($Q_F$) and backward ($Q_B$) hemisphere 
charges, where the forward
hemisphere is defined as the one that contains the momentum vector of the 
incoming electron.

In order to ensure that each event is used at most once in 
this analysis and the measurement based on prompt leptons and slow pions
described below, 
every event is assigned a figure of merit $\Psigvtx$ defined as
\begin{equation}
\Psigvtx = \tilde{F}_{\rm b} \cdot (2 \tilde{P}_{\rm b} -1) \,,
\end{equation}
where $\tilde{F}_{\rm b}$ denotes the estimated $\bb$ purity as a function of 
decay length significance
and $\tilde{P}_{\rm b}$ stands for the estimated probability of correct
charge identification\footnote{The charge identification probability
$\tilde{P}_{\rm b}$ does not depend significantly
on the decay length significance of the tagged vertex.} 
as a function of $|Q_F-Q_B|$, both determined from the simulation.
Events are rejected from the vertex-tagged sample if they are 
also tagged by the presence of a prompt lepton or slow pion with a
corresponding figure of merit $\Psiglsp > \Psigvtx$, where
$\Psiglsp$ is determined according to Equation~\ref{Psig.eqn} 
(see Section~\ref{flavoursep.subsec} below).
At $\sqrt{s}=189\ \GeV$, 
the events that contain both a tagged secondary vertex and a prompt lepton
or slow pion correspond to $56\%$ of the vertex tagged and $28\%$
of the lepton or slow pion tagged samples, respectively.
Of these common events, $58\%$ are assigned to the secondary vertex tagged event
sample by the procedure described above.  It has been checked that
systematic cross-dependences between the two asymmetry measurements, 
which may in principle be introduced by this method, are negligibly small.
Note that the quantity $\Psigvtx$ is not used in the
fit which determines $\AFBb$, but 
only to define the selected sample of events.

For the final vertex-tagged event sample,
the tagging efficiencies for each 
flavour are determined from 
Monte Carlo and 
are shown in Table~\ref{tab:effprob}.
The uncertainties include the systematic errors, 
which will be discussed in detail in
Section~\ref{sec:sys}.
They are dominated by uncertainties in the 
bottom and charm physics modelling and
detector resolution, as in the $\Rb$ analysis.
The b purity is about 60$\%$, 
with a fraction of four-fermion background 
up to 5$\%$. The lower b purity with respect to the folded tag purity does not 
limit the precision of the asymmetry measurement with the present statistics.
Because of the rejection of some of the events that are also lepton or
slow pion tagged, the final $\bb$ efficiencies and purities are lower than 
those which could otherwise be obtained for a vertex-tagged event sample.

\begin{table}[htb]
\begin{center}
$\begin{array}{|c||c|c|}
 \hline
 {\rm Flavour\ q}  & 
 \begin{array}{c}
    {\rm Event\ selection}\\
    {\rm efficiency}\ \epsilon_{\Pq}
 \end{array} &
 \begin{array}{c}
    {\rm Charge\ identification}\\
    {\rm probability}\ P_{\Pq}
 \end{array} \\
\hline
\hline
{\rm d}   & 0.032\pm 0.005 &  0.673 \pm 0.021 \\
{\rm u}   & 0.036\pm 0.005 &  0.743 \pm 0.015 \\
{\rm s}   & 0.035\pm 0.005 &  0.683 \pm 0.020 \\
{\rm c}   & 0.097\pm 0.005 &  0.645 \pm 0.011 \\
{\rm b}   & 0.383\pm 0.025 &  0.683 \pm 0.012 \\
\hline
\end{array}$
\caption{\label{tab:effprob}
Event tagging efficiencies $\epsilon_{\rm q}$ and charge identification
probabilities $P_{\rm q}$ at $\sqrt{s} = 189\ \GeV$, 
as determined from Monte Carlo simulation. 
The efficiencies and charge identification probabilities
are computed for the final selected samples after rejection of events
with $\Psiglsp>\Psigvtx$.  Similar values have been obtained for the
other centre-of-mass energies.}
\end{center}
\end{table}

The quantities $P_{\rm q}$ are the probabilities for 
the hemisphere charge method to correctly identify 
the event hemisphere into which the primary quark ${\rm q}$ was emitted.
They are determined from Monte Carlo simulation and are
given in Table~\ref{tab:effprob} for the different quark flavours
at $\sqrt{s} = 189$~GeV.
Similar values are obtained at different centre-of-mass energies.
Their errors include systematic uncertainties which will be discussed in
detail in Section~\ref{sec:sys}.
The largest contribution to the systematic error of the 
b and c quark charge identification probabilities arises from the
modelling of heavy flavour fragmentation and decay.
For light flavours the uncertainties on fragmentation
are expected to have a small effect on the total systematic error, and
only Monte Carlo statistics 
and detector resolution are considered.
Possible detector biases in the charge identification probability
for positive and negative quarks have been investigated 
using $\PZz$ calibration data, and have been found to be negligible. 

\begin{figure}
\begin{center}
\hspace{85pt}{\Large\bf OPAL}
\end{center}
\vspace{-89pt}
\begin{center}
\epsfig{file=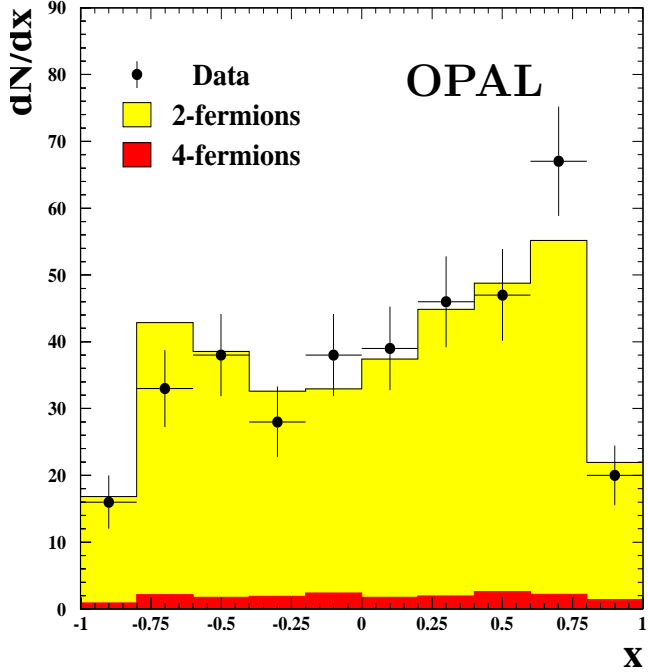,
width=0.55\textwidth,height=10.cm} 
\end{center}
\vspace{-0.5cm}
\caption
{The distribution of the observed $x= -{\rm sign}(Q_F - Q_B)\cdot 
|\cos\theta_T|$ at 189 GeV centre-of-mass energy.
The points with error bars are the data, 
and the histogram is the Monte Carlo expectation.  
The dark shaded histogram represents
the expected contribution from four-fermion
background.  The histogram showing the
Monte Carlo expectation has been scaled to the 
same number of entries as found in the data.}
\label{fig:aobs}
\end{figure}

The quantity 
\begin{equation}
\label{xdef.eqn}
x=-{\rm sign}(Q_F \, - \, Q_B)\cdot | \cos\theta_{T} |
\end{equation}
is computed for each event, where $\theta_{T}$ denotes the polar angle of the
thrust axis.
Its observed distribution at 189~GeV centre-of-mass energy
is compared with the Monte Carlo prediction
in Figure~\ref{fig:aobs}.
When only vector and axial vector couplings of the quarks to a
gauge boson exchanged in the s-channel are allowed, the observed
angular distribution of the primary quark can be expressed 
as~\cite{bib-ZPhysicsatLEP1}
\begin{equation}
\frac{{\rm d}\sigma^{obs}}{{\rm d}x} 
= {\cal C} \, \epsilon (x) \, (1+ x^2 + \frac{8}{3} A^{obs}_{\mathrm{FB}} x)
\,,
\label{eq:asymm}
\end{equation}
where the quark masses have been neglected.
The constant ${\cal C}$ is for normalization, and
$\epsilon(x)$ is the tagging efficiency as a function
of $\cos \theta_T$ for an event.
It is assumed that the efficiencies are symmetric functions of $x$, and it has
been checked in the simulation that their dependence on $x$ is the same
for all primary flavours.
For other event types (e.g.~four-fermion events), the predicted 
differential cross-section is not a second-order polynomial, but
the resulting effects are negligible within the precision of the 
measurements presented here.

Using Equation~\ref{eq:asymm}, the
observed asymmetry $A^{obs}_{\mathrm{FB}}$ is obtained by maximising
the log likelihood
\begin{equation}
\ln {\cal L} =  \sum_{j=1}^N \ln \left\{ {\cal C} \epsilon (x_j) \right\} +
\sum_{j=1}^N \ln \left\{ 1 + x^2_j + \frac{8}{3} A^{obs}_{\mathrm{FB}} x_j \right\}
\,,
\label{eq:likelihood}
\end{equation}
where the sum is over all $N$ selected events, and $A^{obs}_{\mathrm{FB}}$
is the only
free parameter in the fit.
The first term is a constant for a given set of events.

After four-fermion background subtraction, $\AFBb$ is determined 
from the relation
\begin{equation}
 A^{obs}_{\rm FB}=\sum_{\rm q=u,d,s,c,b}s_{\rm q}\cdot F_{\rm q} \cdot 
(2P_{\rm q}-1)\cdot A^{\Pq}_{\rm FB} \ .
\nonumber
\end{equation}
Here, $s_{\rm q}$ is $-1$ ($+1$) for up-type (down-type) quarks, and
$A^{\rm q}_{\rm FB}$ is the forward-backward asymmetry for flavour ${\rm q}$.
The asymmetries for non-b events are fixed to their Standard Model values as 
calculated by ZFITTER. 
The fractions $F_{\rm q}$ of events of flavour ${\rm q}$ in the sample 
are determined as
\begin{equation}
F_{\rm q} = \frac{R_{\rm q} \epsilon_{\rm q}}{\sum_j R_j \epsilon_j} \ ,
\nonumber
\end{equation}
where $R_{\Pq}$ is the ratio  of the cross-section of quark type $\rm q$
to the total hadronic cross-section, 
determined from ZFITTER,
and $\epsilon_{\rm q}$ is the  tagging efficiency determined from Monte Carlo.
The factor $(2P_{\Pq}-1)$ is to account for charge misassignment. 
The small contamination from four-fermion events is
evaluated as for the $\Rb$ measurement and subtracted from the sample.
Its observed asymmetry is found to be consistent with 
zero within the available Monte Carlo statistics.

The numbers of tagged events, the observed asymmetries $A^{obs}_{\rm FB}$,
and the corrected asymmetries $\AFBb$ with their statistical and 
systematic errors are given in 
Table~\ref{tab:afb}, together with the Standard Model expectations at the 
different centre-of-mass energies.
Most of the systematic errors are in common with the $\Rb$
analysis, as discussed below in Section~\ref{sec:sys}. 
The largest systematic errors arise from 
uncertainties in the detector resolution and the event selection procedure.
Uncertainties in the fragmentation of light quarks are assumed to be 
negligible and are not considered. For all centre-of-mass energies, 
the statistical error is dominant.

While the observed asymmetries are well within the physical range
of $-0.75<A^{obs}_{\rm FB}<0.75$, values of $\AFBb$ outside this range
are possible because of the corrections due to sample composition
and charge identification probability.
All corrected $\AFBb$ values are compatible with a value inside
the physical range.
No constraint is applied to force the corrected $\AFBb$ values to 
lie within this range, in order to facilitate the combination with the
values determined in the lepton and slow pion analysis and with
measurements by other experiments.

\begin{table}[htb]
\begin{center}
\begin{tabular}{|c||c||l|r@{\,}c@{\,}l||c|}
\hline
  Energy 
& 
  Events 
& 
 \multicolumn{1}{c|}{$A^{obs}_{\rm FB}$}
& 
 \multicolumn{3}{c||}{$A^{\rm b}_{\rm FB}$}
& 
 $A^{\rm b,\,SM}_{\rm FB}$\rule[-2ex]{0pt}{5.5ex}\\
\hline
\hline
  133 GeV 
 & 
  \enspace 75
 &
  $\phantom{-} 0.12 \pm 0.12$
 & 
  $\phantom{-} 0.67$&$\pm 0.50$&$\pm0.06$\rule[-2ex]{0pt}{5.5ex}
 &
  $0.48$ 
 \\ 
\hline
  161 GeV 
 & 
  \enspace 35
 &
  $-0.17 \enspace \ ^{+0.18\phantom{0}}_{-0.16\phantom{0}}$
 & 
  $-0.59$&$^{+0.76}_{-0.68}$&$\pm0.04$\rule[-2ex]{0pt}{5.5ex}
 & 
  $0.55$  
 \\
\hline
  172 GeV 
 & 
  \enspace 14
 &
  $\phantom{-} 0.32 \enspace \ ^{+0.27\phantom{0}}_{-0.25\phantom{0}}$
 & 
  $\phantom{-} 1.9 \enspace$&$\pm 1.3 \enspace$&$\pm0.1\enspace $\rule[-2ex]{0pt}{5.5ex}
 & 
  $0.56$  
 \\
\hline
  183 GeV 
 & 
  157
 &
  $\phantom{-} 0.182 \ ^{+0.082} _{-0.073}$  
 & 
  $\phantom{-} 1.15$&$^{+0.41}_{-0.37}$&$\pm0.08$\rule[-2ex]{0pt}{5.5ex} 
 & 
  $0.57$  
 \\
\hline
  189 GeV 
 & 
  372
 &
  $\phantom{-} 0.194 \ ^{+0.055}_{-0.047}$
 & 
  $\phantom{-} 1.22 $&$^{+0.28}_{-0.24}$&$\pm0.08$\rule[-2ex]{0pt}{5.5ex}
 &
  0.58
 \\
\hline
\end{tabular}
\caption{\label{tab:afb} 
The numbers of tagged events with the vertex 
tag analysis, the observed asymmetries, the resulting
$\AFBb$ values, and their 
Standard Model predictions.
The first error on $\AFBb$ is statistical, the second 
systematic. 
}
\end{center}
\end{table}


\subsection{Identification of Leptons and Slow Pions}
\label{leptid.subsec}

Prompt leptons 
from semileptonic decays of heavy hadrons provide a means of tagging
both $\bb$ and $\cc$ events that is largely 
independent of the secondary vertex tag.  In addition, slow pions 
from $\PDstp\to\PDz\Ppip$ decays are used for tagging
heavy flavour events. Both prompt leptons and slow pions allow 
a clean identification of the
event hemisphere that contains the primary quark.

\subsubsection{Electron Identification}

Electron candidates are required to have a momentum of at least $2\ \GeV$.
Tracks with less than 20 d$E$/d$x$ samplings in the tracking chamber
are rejected to ensure
a good measurement of the specific energy loss.
The difference between the measured energy loss 
and that expected for an electron, divided by the measurement error,
is required to
be between $-2$ and $4$.

In this sample of preselected tracks, 
electrons are identified with the help of an artificial 
neural network,
which is described in detail in~\cite{bib-Rb}.
In addition to the electron preselection, a network output $\netel>0.9$
is required.  
At $189\ \GeV$, this selection has an
efficiency for prompt electrons of approximately $25\%$,
defined with respect to the total number of prompt electrons that are 
reconstructed as tracks in the detector.  The resulting sample
is $75\%$ pure in electrons.

\begin{figure}[p]
  \begin{center}
  \unitlength 1pt
  \vspace{30pt}
  \hspace{6pt}
  \begin{picture}(500,740)
    \put(26,248){\epsfig
{file=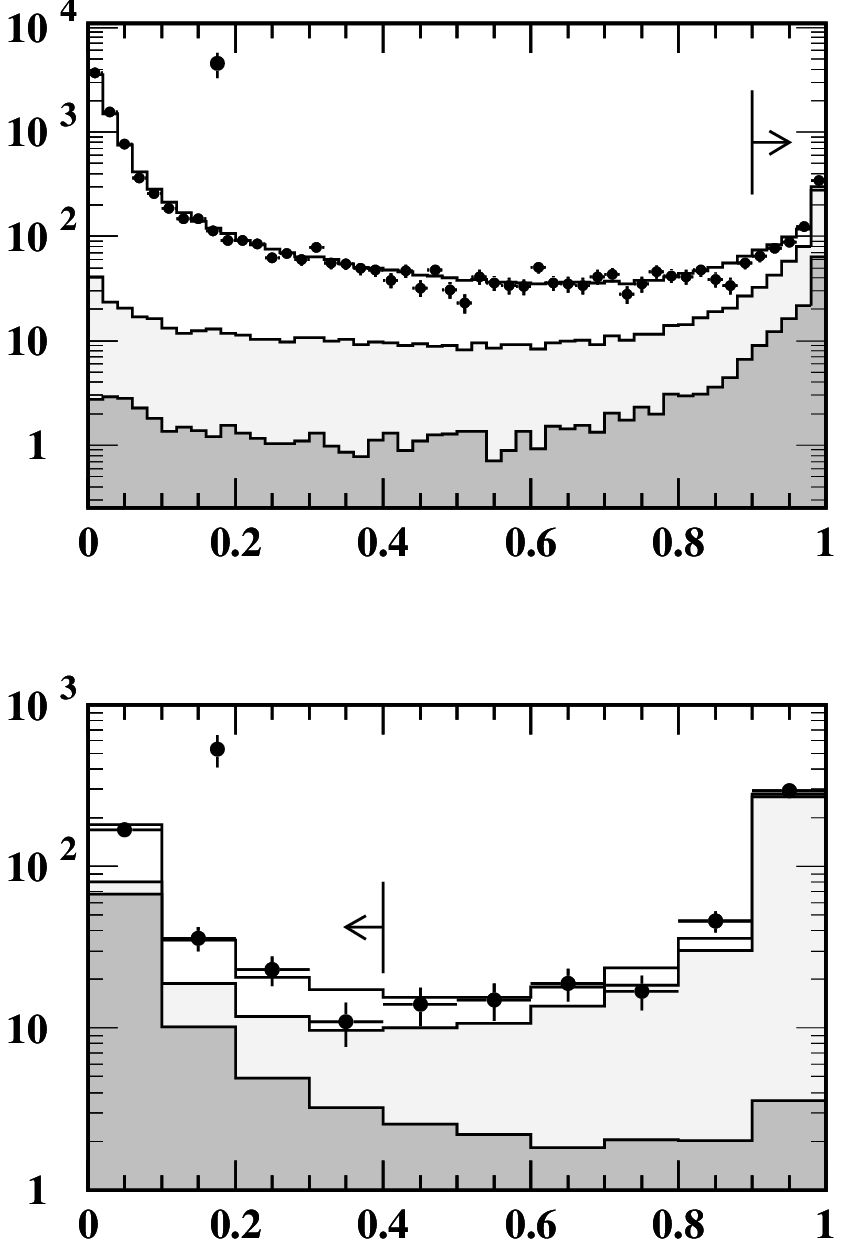,width=730pt,bbllx=0pt,bblly=405pt,bburx=550pt,bbury=792pt}}
    \put(163,738){\bf data}
    \put(126,613){\epsfig
{file=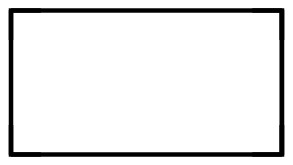,width=120pt,height=120pt,bbllx=0pt,bblly=405pt,bburx=550pt,bbury=792pt}}
    \put(163,722){\bf background}
    \put(126,597){\epsfig
{file=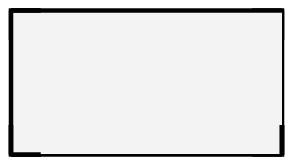,width=120pt,height=120pt,bbllx=0pt,bblly=405pt,bburx=550pt,bbury=792pt}}
    \put(126,597){\epsfig
{file=box.eps,width=120pt,height=120pt,bbllx=0pt,bblly=405pt,bburx=550pt,bbury=792pt}}
    \put(163,706){\bf conversion electrons}
    \put(126,581){\epsfig
{file=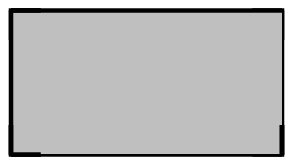,width=120pt,height=120pt,bbllx=0pt,bblly=405pt,bburx=550pt,bbury=792pt}}
    \put(126,581){\epsfig
{file=box.eps,width=120pt,height=120pt,bbllx=0pt,bblly=405pt,bburx=550pt,bbury=792pt}}
    \put(163,689){\bf prompt electrons}
    \put(163,475){\bf data}
    \put(126,351){\epsfig
{file=box.eps,width=120pt,height=120pt,bbllx=0pt,bblly=405pt,bburx=550pt,bbury=792pt}}
    \put(163,459){\bf background}
    \put(126,335){\epsfig
{file=shadedbox2.eps,width=120pt,height=120pt,bbllx=0pt,bblly=405pt,bburx=550pt,bbury=792pt}}
    \put(126,335){\epsfig
{file=box.eps,width=120pt,height=120pt,bbllx=0pt,bblly=405pt,bburx=550pt,bbury=792pt}}
    \put(163,443){\bf conversion electrons}
    \put(126,319){\epsfig
{file=shadedbox7.eps,width=120pt,height=120pt,bbllx=0pt,bblly=405pt,bburx=550pt,bbury=792pt}}
    \put(126,319){\epsfig
{file=box.eps,width=120pt,height=120pt,bbllx=0pt,bblly=405pt,bburx=550pt,bbury=792pt}}
    \put(163,429){\bf prompt electrons}
    \put(360.5,533.0){\Large\boldmath${\cal N}_{\rm el}$}
    \put(356.5,272.0){\Large\boldmath${\cal N}_{\rm cv}$}
    \put(113,730.0){\Large\bf (a)}
    \put(113,470.0){\Large\bf (b)}
    \put(272,723.0){{\huge\bf OPAL}}
    \put(272,461.0){{\huge\bf OPAL}}
    \put(62.0,697.0){\rotninety\rotninety\rotninety\Large\bf Entries}
    \put(58.0,436.0){\rotninety\rotninety\rotninety\Large\bf Entries}
  \end{picture}
  \vspace{-275pt}
  \caption{\label{nnel.fig}
The $\netel$ distribution is shown in figure (a) for tracks 
at $189\ \GeV$ centre-of-mass energy 
that pass the electron momentum and d$E$/d$x$ cuts.
In (b), the $\netcv$ distribution is given for those tracks that
also pass the requirement of $\netel>0.9$.
In each case, 
the points with error bars represent the data distribution, and the 
histograms the Monte Carlo simulation, scaled to the same number of entries.
The dark and light grey areas are the expected contributions from 
prompt electrons and conversion electrons, respectively, while the open
area corresponds to hadrons.  The arrows indicate
the accepted regions.
\vspace{10pt}
}
  \end{center}
\end{figure}

After this selection, electrons from photon conversions are 
an important background in the sample. 
A separate artificial neural network is used
to identify pairs of conversion electrons~\cite{bib-OPALlafb,bib-Rb}. 
The contribution from photon 
conversions is reduced by requiring 
a network output of $\netcv < 0.4$.
In the Monte Carlo simulation at $189\ \GeV$ centre-of-mass energy,
$89\%$ of the electrons from photon conversions are rejected by this cut, 
while $90\%$ of the prompt electrons are kept.
In Figure~\ref{nnel.fig}, the $\netel$ and $\netcv$ output distributions are
shown for tracks at $189\ \GeV$ centre-of-mass energy.  

\subsubsection{Muon Identification}

The muon selection proceeds in two steps.  First, muon track segments are
formed from the hits in the muon chambers.  Tracks from the central tracking 
chambers with a momentum greater than $2\ \GeV$ are extrapolated to the 
muon chambers.  For each track segment in the muon chambers, only the ``best 
matching track'' is considered for use in the asymmetry fit.  It
is defined as the extrapolated 
track that has the smallest angular separation~$\alpha$ to the muon track 
segment in question.

\begin{figure}[htb]
  \begin{center}
  \unitlength 1pt
  \vspace{-261pt}
  \begin{picture}(450,492)
    \put(26,-274){\epsfig
{file=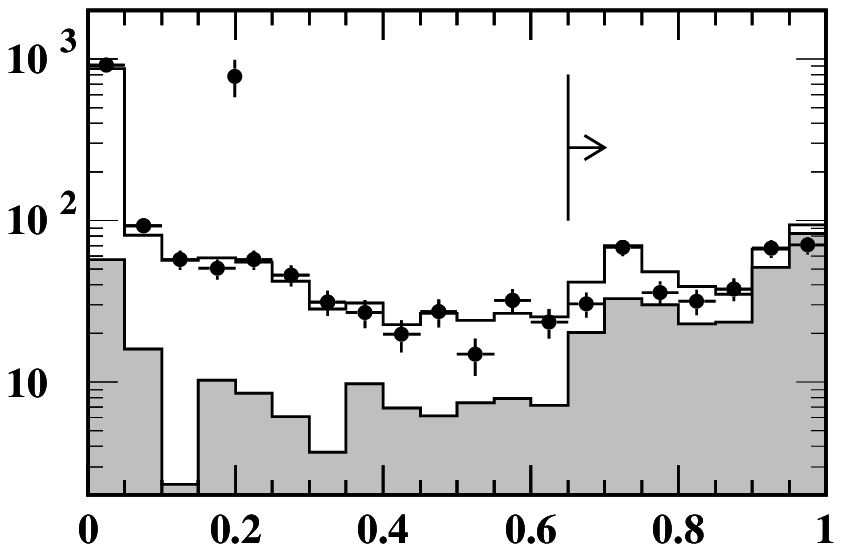,width=700pt,bbllx=0pt,bblly=405pt,bburx=550pt,bbury=792pt}}
    \put(163,185){\bf data}
    \put(126,60){\epsfig
{file=box.eps,width=120pt,height=120pt,bbllx=0pt,bblly=405pt,bburx=550pt,bbury=792pt}}
    \put(163,169){\bf background}
    \put(126,44){\epsfig
{file=shadedbox7.eps,width=120pt,height=120pt,bbllx=0pt,bblly=405pt,bburx=550pt,bbury=792pt}}
    \put(126,44){\epsfig
{file=box.eps,width=120pt,height=120pt,bbllx=0pt,bblly=405pt,bburx=550pt,bbury=792pt}}
    \put(163,153){\bf muons}
    \put(347.5,-3.0){\Large\boldmath${\cal N}_\mu$}
    \put(279.0,182.0){{\huge\bf OPAL}}
    \put(60.0,155.0){\rotninety\rotninety\rotninety\Large\bf Entries}
  \end{picture}
  \vspace{10pt}
  \caption{\label{nnmu.fig}
The $\netmu$ distribution for best matching tracks 
at $189\ \GeV$ centre-of-mass energy 
that pass the muon momentum cut.
The points with error bars represent the data distribution, and the 
histogram the Monte Carlo simulation, scaled to the same number of entries.
The shaded area is the expected contribution from true muons.
The accepted region is indicated by the arrow.
}
  \end{center}
\end{figure}

In a second step, an artificial 
neural network trained for muon identification is used
to enhance the purity of the muon sample.  
The network uses the following eleven inputs:
\begin{list}{$\bullet$}{\setlength{\itemsep}{0ex}
                        \setlength{\parsep}{1ex}
                        \setlength{\topsep}{0ex}}
\item Information from the matching:
  \begin{list}{$-$}{\setlength{\itemsep}{0ex}
                    \setlength{\parsep}{0ex}
                    \setlength{\topsep}{0ex}}
  \item
    The square root of the $\chi^2$ for the position match in $\theta$ 
    and $\phi$ between the extrapolated track and the associated muon 
    track segment in the muon chambers, as described in~\cite{bib-muonsel};
  \item
    the ratio of distances $\Rmis = \alpha^{(1)}/\alpha^{(2)}$ of the best and
    second best matching track to the muon segment;
    this is a measure of how ambiguous the choice of the best matching track
    was in the preselection; 
  \item
    the $\chi^2$ probability for the matching computed using
    both position and direction information for the track in the central
    detector and the associated muon track segment.
  \end{list}
\item Information from the hadron calorimeter:
  \begin{list}{$-$}{\setlength{\itemsep}{0ex}
                    \setlength{\parsep}{0ex}
                    \setlength{\topsep}{0ex}}
  \item
    The number of calorimeter layers in the cluster associated with the central
    track;
  \item
    the number of the outermost such layer; 
  \item
    the $\chi^2$ probability for the match in $\theta$ and $\phi$ between 
    the track (extrapolated to the hadron calorimeter) and the associated
    cluster.
  \end{list}
\item Specific energy loss:
  \begin{list}{$-$}{\setlength{\itemsep}{0ex}
                    \setlength{\parsep}{0ex}
                    \setlength{\topsep}{0ex}}
  \item
    The muon $\dEdx$ weight for the track, which is a measure of the probability that
    the track is compatible with a muon hypothesis;
  \item
    $\sigma_{{\rm d}E/{\rm d}x}$, the error on the $\dEdx$ measurement;  
  \item
    the momentum of the track.
  \end{list}
\item Geometrical information:
  \begin{list}{$-$}{\setlength{\itemsep}{0ex}
                    \setlength{\parsep}{0ex}
                    \setlength{\topsep}{0ex}}
  \item
    The position in $|\cos\theta|$ and $\phi$ where the extrapolated track
    enters the muon chambers.
  \end{list}
\end{list}
The distribution of the neural network output $\netmu$ is shown in 
Figure~\ref{nnmu.fig} for ``best matching'' tracks according to the 
definition above. Muon candidates are retained if $\netmu$ is larger than 
$0.65$.  In Monte Carlo simulated events at 189 GeV centre-of-mass energy, the
muon selection 
results in an efficiency of $43\%$ for prompt muons,
defined with respect to all prompt muons that are reconstructed as tracks,
and a muon purity of $73\%$.

\subsubsection{Preselection of Slow Pion Candidates}
\label{spsel.subsec}

Pions from $\PDstp\to\PDz\Ppip$ decays, denoted $\spi$ in the following,
are selected based on the kinematic
properties of this decay.  Due to the low momentum, 
$p^*=39\ \MeV$~\cite{bib-PDG1998}, of the decay products in the $\PDstp$ rest
frame, pions from this decay have momenta smaller than
\begin{equation}
p_{\pi_s}^{\rm max} = \frac{\sqrt{s}}{2 m_{\PDstp}}
                      \left(E^* + p^*\right) 
                    = 0.0458\,\sqrt{s}
\end{equation}
in the laboratory frame, where $E^* = \sqrt{(p^*)^2 + m_{\Ppip}^2}$, and
$m_{\PDstp}$ and $m_{\Ppip}$ denote the $\PDstp$ and $\Ppip$ masses, 
respectively.
In addition, slow pions have a transverse momentum with respect to the 
$\PDstp$ flight direction of at most $p^*$, and are thus dominantly found
in the core of the jet containing the $\PDstp$ meson.

\begin{figure}[htb]
  \begin{center}
  \unitlength 1pt
  \vspace{-261pt}
  \begin{picture}(450,492)
    \put(26,-274){\epsfig
{file=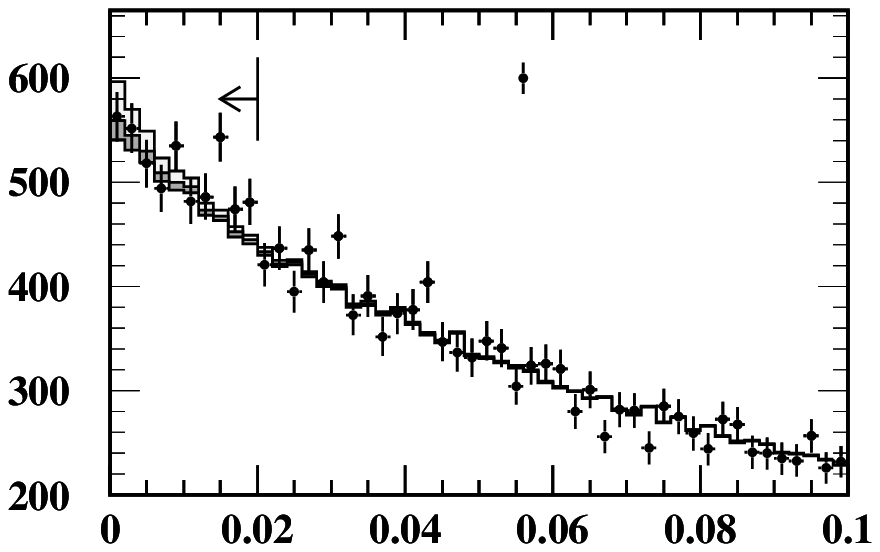,width=700pt,bbllx=0pt,bblly=405pt,bburx=550pt,bbury=792pt}}
    \put(263,185){\bf data}
    \put(226,60){\epsfig
{file=shadedbox2.eps,width=120pt,height=120pt,bbllx=0pt,bblly=405pt,bburx=550pt,bbury=792pt}}
    \put(226,60){\epsfig
{file=box.eps,width=120pt,height=120pt,bbllx=0pt,bblly=405pt,bburx=550pt,bbury=792pt}}
    \put(263,169){\boldmath${{\rm c}\!\to\!\spi}$}
    \put(226,44){\epsfig
{file=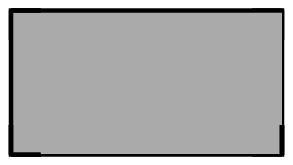,width=120pt,height=120pt,bbllx=0pt,bblly=405pt,bburx=550pt,bbury=792pt}}
    \put(226,44){\epsfig
{file=box.eps,width=120pt,height=120pt,bbllx=0pt,bblly=405pt,bburx=550pt,bbury=792pt}}
    \put(263,153){\boldmath${{\rm b}\!\to\![{\rm c},\overline{\rm c}]\!\to\!\spi}$}
    \put(226,28){\epsfig
{file=box.eps,width=120pt,height=120pt,bbllx=0pt,bblly=405pt,bburx=550pt,bbury=792pt}}
    \put(263,137){\bf background}
    \put(279.5,-3.0){\Large\bf \boldmath$\ptsq$\unboldmath\ in GeV\boldmath$ ^2$\unboldmath}
    \put(121.5,52.0){{\huge\bf OPAL}}
    \put(49.0,155.0){\rotninety\rotninety\rotninety\Large\bf Entries}
  \end{picture}
  \vspace{10pt}
  \caption{\label{pt2.fig}
The $\ptsq$ distribution for tracks at $189\ \GeV$
centre-of-mass energy that pass the slow pion momentum and $\protect{\dEdx}$ 
cuts.  Tracks that form single charged particle jets have been excluded.
The points with error bars represent the data distribution, and the 
histogram the Monte Carlo simulation, scaled to the same number of entries.
The dark grey area is the expected contribution from true slow pions
from cascade $\protect{\btoccbartosp}$ decays, and the light grey area 
represents $\protect{\ctosp}$ decays.
The arrow indicates the accepted region.
}
  \end{center}
\end{figure}

Slow pion candidates are required to have a momentum between
$1.0\ {\rm GeV}$ and $p_{\pi_s}^{\rm max}$
and, if at least 20 $\dEdx$ samplings are available,
a specific energy loss $\dEdx$ whose probability compatibility for the
pion hypothesis 
exceeds $2\%$.
Tracks that form single charged particle jets are rejected.
The $\PDstp$ flight direction is estimated by the jet direction, which
is recalculated in an iterative procedure similar to the one described 
in~\cite{bib-Ties}, based on the rapidities of the tracks and clusters in 
the jet containing the slow pion candidate.
If the jet mass exceeds $2.3\ \GeV$, the track or calorimeter cluster in the 
jet with the smallest rapidity with respect to the jet axis
is removed from the calculation, and the direction is recomputed. 
The transverse momentum $\pt$ 
of the slow pion candidate is calculated with respect to this jet direction.
The $\ptsq$ distribution is shown in Figure~\ref{pt2.fig} for 
tracks at $189\ \GeV$ centre-of-mass energy.  
Slow pion candidates are
accepted if $\ptsq < 0.02\ \GeV^2$.

In Monte Carlo simulated events at $189\ \GeV$, 
this preselection is 
$56\%$ efficient and yields a sample that contains $5.7\%$ of slow pions.
This sample is further enriched with a cut that is described in 
Section~\ref{flavoursep.subsec} below.

\subsection{Flavour Separation of the Lepton and Slow Pion Samples}
\label{flavoursep.subsec}

Three different sources of prompt leptons are considered: 
$\btol$, meaning leptons from semileptonic 
decays of b-flavoured hadrons;
cascade bottom decays, which include the contributions
from $\btoctol$ and $\btocbartol$ processes; and $\ctol$, leptons from 
semileptonic decays of charm hadrons. 
The background can be classified as ``non-prompt'' leptons, i.e.~all 
other leptons that are not produced in the decay of b- or c-flavoured hadrons, 
and particles that are mis-identified as electrons or muons. 

For electrons and muons, separate artificial 
neural networks have been constructed with the aim of separating 
prompt $\btol$ decays, cascade $\btoccbartol$ decays, prompt $\ctol$ 
decays, and all other contributions.
The technique used is similar to the one described in~\cite{bib-OPALlafb}, 
but the inputs and training have been re-optimized for
centre-of-mass energies above the $\PZz$ resonance; also an artificial 
neural network for
the identification of cascade decays is included.
Candidates that form single charged particle jets are rejected since they
are expected to be dominantly produced in leptonic W decays.

The first network, denoted $\netb$, is 
designed to separate $\btol$ decays from all other contributions.
Two networks, $\netbc$ and $\netc$, have been trained on a Monte
Carlo simulation that does not contain $\btol$ decays in order to classify
lepton candidates that are background to the $\netb$ net.
All three networks use the following input variables, where jet variables are 
defined with the lepton candidate included in the jet:
\begin{list}{$\bullet$}{\setlength{\itemsep}{0ex}
                        \setlength{\parsep}{0ex}
                        \setlength{\topsep}{0ex}}
\item 
$p$, the lepton track momentum;
\item 
$p_t$, the transverse momentum of the lepton track calculated relative 
to the jet which contains the track;
\item 
$L/\sigma_{L}$, the 
decay length significances of the secondary vertices (if existing)
in the jet containing the lepton
and the most energetic jet in the hemisphere not containing the lepton,
where secondary vertices are reconstructed with the same algorithm as 
described in Section~\ref{sec:vtx};
\item 
the jet charges of the jets containing 
the lepton and the most energetic jet in the hemisphere not containing 
the lepton, each multiplied by the lepton charge, where
the jet charge is defined as in Equation~\ref{defjq} with $\kappa=0.4$, 
but using only tracks associated to the jet;
\item 
the forward multiplicity in the lepton jet, defined as the number
of tracks with an impact parameter significance with respect to the 
primary vertex larger than $2$.  For each track, 
the impact parameter is defined as the distance 
between the primary vertex and the track at its the point of closest
approach; the impact parameter significance is defined as this distance
divided by its error;
\item 
the $|\cos\theta|$ of the jet momentum vector, where $\theta$ is the jet
polar angle;
\item 
the outputs $\netel$ and $\netcv$ of the electron 
identification network and the conversion finder network, 
respectively, in the case of electrons;  
\item 
the output $\netmu$ of the muon
identification network, in the case of muons.
\end{list}

From the $\netb$, $\netbc$, and $\netc$ network outputs,
the following quantities are computed which are 
related to the probabilities of a lepton candidate to come from 
one of the three sources:
\begin{eqnarray}
\label{Psig0.eqn}
\nonumber
  \Psigbl  & = & \protect{\netb} 
                 \\
  \Psigbcl & = & \left( 1 - \netb \right)
                \frac{ \left[ \netbc \left( 1 - \netc \right) \right]}
                     {   \left[ \netbc \left( 1 - \netc  \right) \right]
                       + \left[ \netc  \left( 1 - \netbc \right) \right]
                       + \left[ \left( 1 - \netc  \right) 
                                \left( 1 - \netbc \right) \right] } 
                \\
\nonumber
  \Psigcl  & = & \left( 1 - \netb \right)
                \frac{ \left[ \netc \left( 1 - \netbc \right) \right]}
                     {   \left[ \netbc \left( 1 - \netc  \right) \right]
                       + \left[ \netc  \left( 1 - \netbc \right) \right]
                       + \left[ \left( 1 - \netc  \right) 
                                \left( 1 - \netbc \right) \right] } 
                \ .
\end{eqnarray}
The difference in the 
treatment of the $\netb$ output from $\netbc$ and $\netc$ is
due to the fact that $\btol$ decays were omitted in the training of the 
latter two networks. Only candidates that satisfy the condition
\begin{equation}
\label{Psig.eqn}
  \Psigl = \sqrt{  \left(\Psigbl\right)^2 + \left(\Psigbcl\right)^2 
                 + \left(\Psigcl\right)^2} > 0.1
\end{equation}
are used in the subsequent analysis.  Candidates with lower values of $\Psigl$
are expected to be dominantly background and to 
have a negligible contribution to the overall result.
Note that the quantity $\Psigl$, although used to define the selected sample
of events, is not itself used in the fit that determines the bottom
and charm asymmetries.

Similarly to the lepton case, slow pion candidates are classified as cascade 
$\btoccbartosp$ decays, $\ctosp$, and background.  Another two 
artificial neural networks, $\netbc$ and $\netc$, have been trained for the
separation of the three components in the preselected slow pion sample.  
The following inputs are used by both networks:
\begin{list}{$\bullet$}{\setlength{\itemsep}{0ex}
                        \setlength{\parsep}{0ex}
                        \setlength{\topsep}{0ex}}
\item 
$p$, the slow pion track momentum;
\item
$\ptsq$, the transverse momentum squared 
of the slow pion track calculated relative 
to the jet direction obtained with the same rapidity based algorithm
that is used in Section~\ref{spsel.subsec};
\item 
$E_{\spi-\rm jet}$, the total energy of the jet containing the slow pion;
\item 
$E_{\spi-{\mathrm sub-jet}}$, the 
energy of the sub-jet~\cite{bib-lsubjet} containing the slow pion:
Each jet containing a slow pion is split into two sub-jets, where the slow pion
sub-jet is seeded by the slow pion track. In an iterative procedure, 
any particle that forms a smaller opening angle with the slow pion sub-jet
than with the remainder of the jet is then assigned to the slow pion sub-jet;
\item 
$L/\sigma_{L}$, 
the 
decay length significance 
of the vertex in the jet containing the slow pion, if a vertex is found;
\item 
the jet charge of the jet containing the slow pion, 
calculated with $\kappa=0.4$, 
multiplied by the slow pion charge; and
\item 
the $|\cos\theta|$ of the jet momentum vector.
\end{list}
In addition, the network for identification of $\btoccbartosp$ decays uses:
\begin{list}{$\bullet$}{\setlength{\itemsep}{0ex}
                        \setlength{\parsep}{0ex}
                        \setlength{\topsep}{0ex}}
\item 
$L$, the decay length of the vertex (if existing) in the jet 
containing the slow pion;
\item 
the jet charge of the jet containing the slow pion,
calculated with a different parameter $\kappa=2.0$,
multiplied by the slow pion charge;
\item 
the forward multiplicity in the jet containing the slow pion, defined as above;
\item 
$(\sum p_t)_{\rm jet}$, the scalar sum of the transverse momenta
relative to the jet axis of all tracks in the jet; and
\item 
the maximum longitudinal momentum component of any track in the 
jet containing the slow pion, measured relative to the jet direction.
\end{list}
The quantities $\Psigbcsp$ and $\Psigcsp$ are computed in analogy to
Equation~\ref{Psig0.eqn} for prompt lepton
candidates, while
$\Psigbsp$ is set to zero.  The same requirement on $\Psigsp$ 
(computed according to Equation~\ref{Psig.eqn}) is imposed as
for prompt leptons. 

If more than one lepton or slow pion candidate per event 
passes the selection, 
the one with the highest $\Psiglsp$ value is taken. 
Furthermore, this $\Psiglsp$ value is required to be larger than the value
of $\Psigvtx$ as determined for the hemisphere charge measurement
(see Section~\ref{sec:afb}) if the event is also tagged by the presence
of a secondary vertex.

\begin{figure}[p]
  \begin{center}
  \unitlength 1pt
  \begin{picture}(450,540)
    \put(-34,174){\epsfig
{file=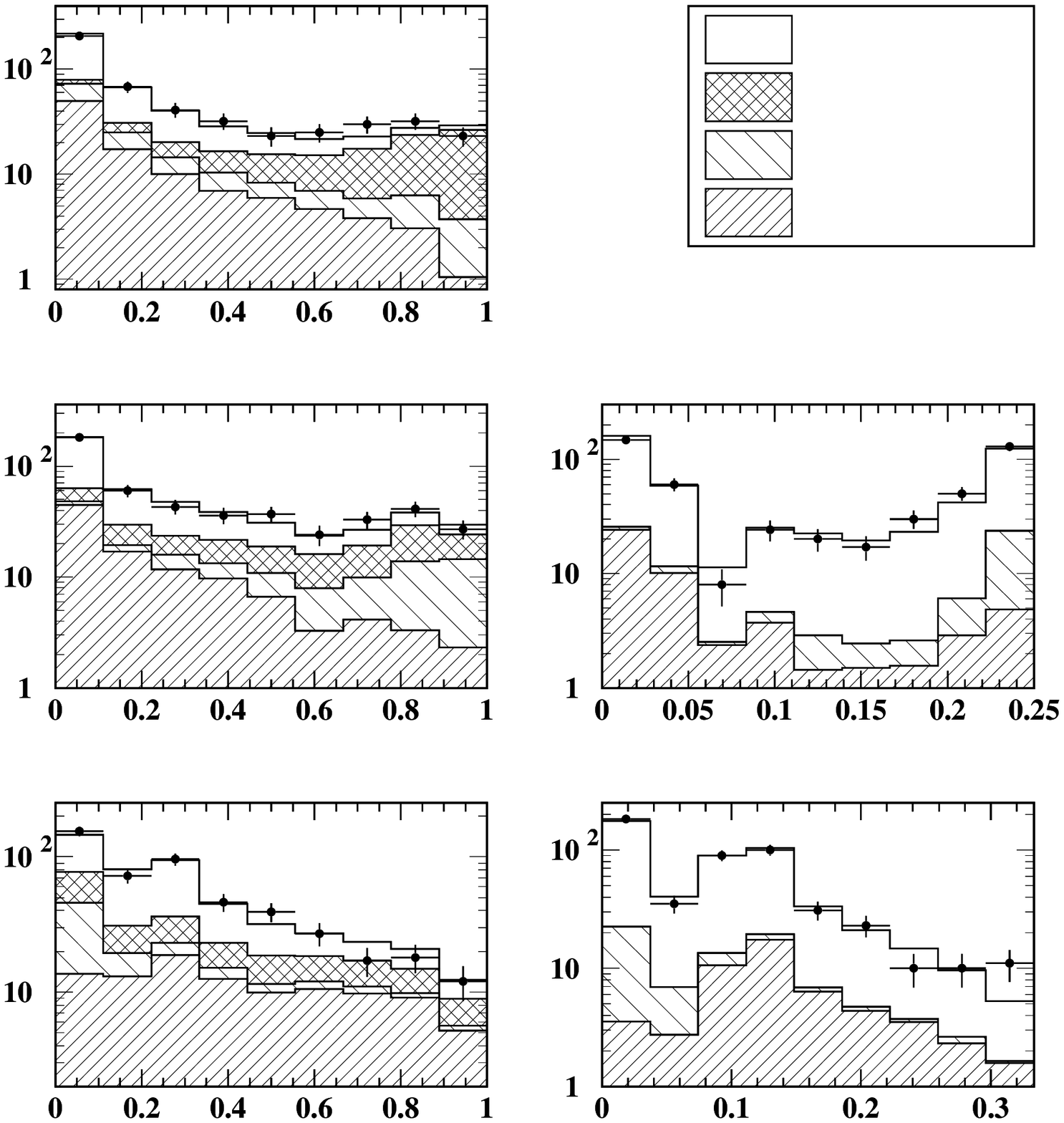,width=500pt,bbllx=0pt,bblly=405pt,bburx=550pt,bbury=792pt}}
    \put(84.0,535.0){\Large\bf Leptons}
    \put(316.0,355.0){\Large\bf Slow Pions}
    \put(355.5,502.0){\bf background}
    \put(355.5,476.0){\boldmath $\btolshort$\unboldmath}
    \put(355.5,450.0){\boldmath $\btoccbartolspshort$\unboldmath}
    \put(355.5,424.0){\boldmath $\ctolspshort$\unboldmath}
    \put(199.0,555.0){\huge\bf OPAL}
    \put(189.0,363.0){\Large\boldmath$\netb$\unboldmath}
    \put(182.0,183.0){\Large\boldmath$\netbc$\unboldmath}
    \put(189.0,3.0){\Large\boldmath$\netc$\unboldmath}
    \put(429.0,183.0){\Large\boldmath$\netbc$\unboldmath}
    \put(436.0,3.0){\Large\boldmath$\netc$\unboldmath}
    \put(-11.5,463.0){\rotninety\rotninety\rotninety\Large\bf Entries}
    \put(-15.0,285.0){\rotninety\rotninety\rotninety\Large\bf Entries}
    \put(-18.5,105.0){\rotninety\rotninety\rotninety\Large\bf Entries}
    \put(224.5,285.0){\rotninety\rotninety\rotninety\Large\bf Entries}
    \put(220.0,105.0){\rotninety\rotninety\rotninety\Large\bf Entries}
  \end{picture}
  \caption{\label{nnbclsp.fig}
The $\netb$, $\netc$, and $\netbc$ distributions for lepton and slow pion
tagged events at $189\ \GeV$ centre-of-mass energy.  
The points with error bars correspond to the
data.  The open histogram represents the Monte Carlo expectation, where
the contributions from signal events are shown hatched as explained in the
figure.
The purity in the slow pion selection is limited, which is reflected in the
fact that the range of $\netc$ and $\netbc$ values for slow pions does
not extend up to~1.
}
  \end{center}
\end{figure}

A breakdown of the composition of the samples of 
electron, muon, and slow pion tagged events at 189 GeV centre-of-mass energy
together with the efficiencies
is given in Table~\ref{tab-lidperf}. 
In Figure~\ref{nnbclsp.fig},
the output distributions of the flavour separation 
networks are shown for the events that pass all cuts.
\begin{table}[htb]
\begin{center}
\begin{tabular}{|c||c|c|c|ccc}
\hline
            & \multicolumn{3}{c}{Sample composition}
            & \multicolumn{3}{||c|}{Efficiencies} \\
Source      & Electrons  & Muons & Slow pions 
            & \multicolumn{1}{||c|}{Electrons} & \multicolumn{1}{c|}{Muons} 
            & \multicolumn{1}{c|}{Slow pions} \\
\hline\hline
$\btol$     & 19\,\% & 19\,\% & ---
            & \multicolumn{1}{||c|}{26\,\%} 
            & \multicolumn{1}{c|}{50\,\%} 
            & \multicolumn{1}{c|}{---} \\
$\btoccbartolsp$
            & \enspace 9\,\% & 11\,\% & \enspace 3\,\%
            & \multicolumn{1}{||c|}{16\,\%} 
            & \multicolumn{1}{c|}{31\,\%} 
            & \multicolumn{1}{c|}{\enspace 6\,\%} \\
$\ctolsp$   & 19\,\% & 24\,\% & 13\,\%
            & \multicolumn{1}{||c|}{18\,\%} 
            & \multicolumn{1}{c|}{38\,\%} 
            & \multicolumn{1}{c|}{13\,\%} \\
\hline
Non-prompt leptons & 15\,\% & 20\,\% & --- & & & \\
Other background     & 37\,\% & 25\,\% & 85\,\% & & & \\
\cline{1-4}
\end{tabular}
\caption{The estimated composition of the lepton and slow pion
tagged event samples at $189\ \GeV$ centre-of-mass energy
after the selection.  The
efficiencies of the prompt lepton and slow pion selection are given
for the final event sample with respect to all prompt leptons and slow pions
that are reconstructed as tracks in the detector.}
\label{tab-lidperf}
\end{center}
\end{table}

\subsection{Measurement of \boldmath$\AFBb$ and $\AFBc$\unboldmath{ }with Leptons
and Slow Pions}
\label{fit.subsec}

\begin{figure}[htb]
  \begin{center}
  \unitlength 1pt
  \begin{picture}(450,540)
    \put(-24,174){\epsfig
{file=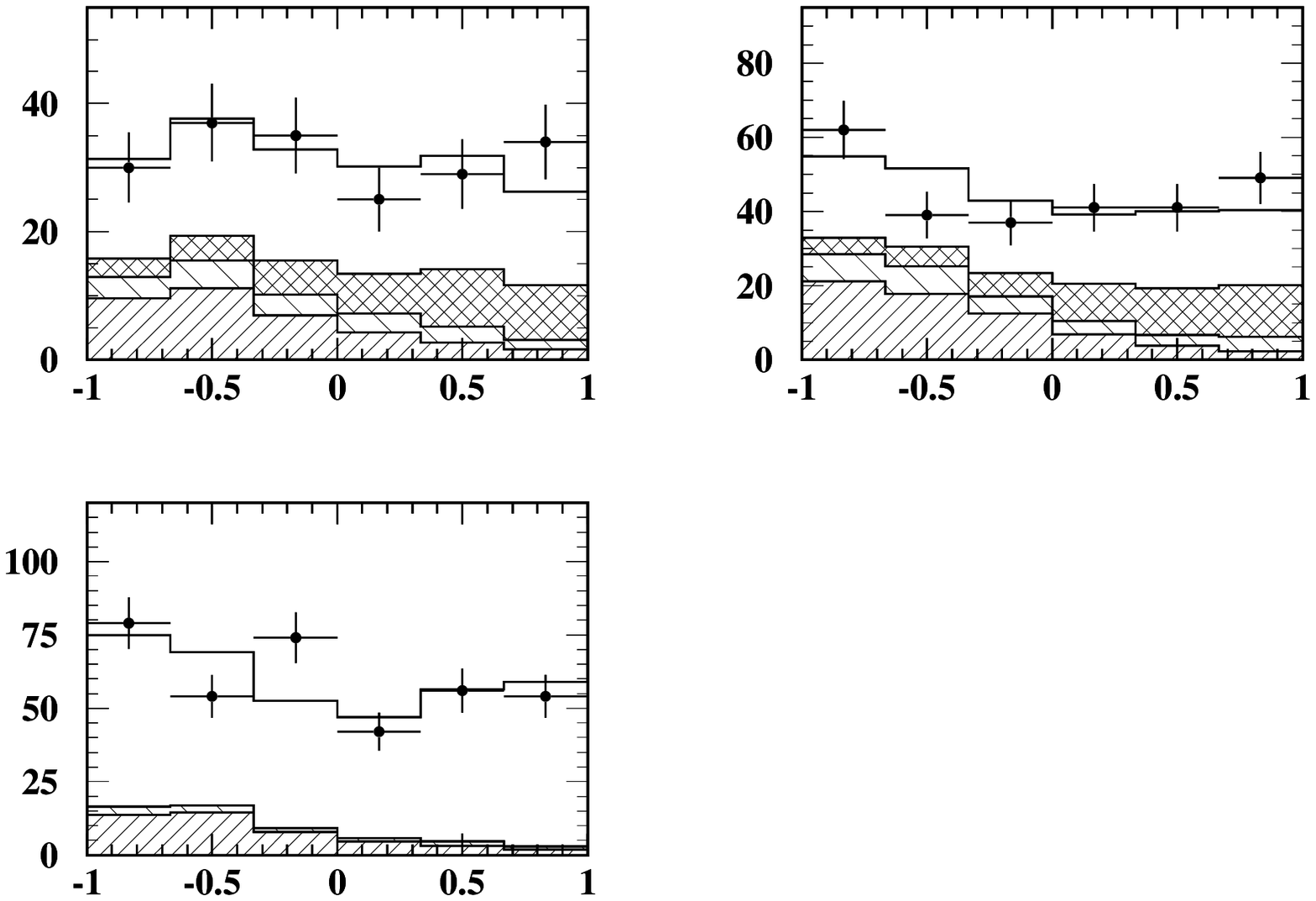,width=500pt,bbllx=0pt,bblly=405pt,bburx=550pt,bbury=792pt}}
    \put(-24,160){\epsfig
{file=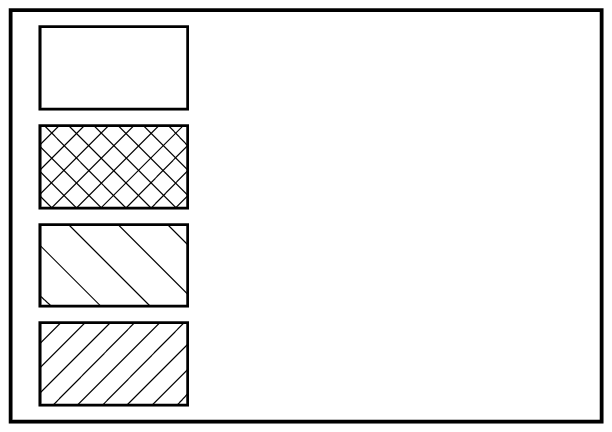,width=500pt,bbllx=0pt,bblly=405pt,bburx=550pt,bbury=792pt}}
    \put(38,499.0){\Large\bf (a) electron}
    \put(296,499.0){\Large\bf (b) muon}
    \put(38,321.0){\Large\bf (c) slow pion}
    \put(187,533.0){{\huge\bf OPAL}}
    \put(365.5,304.0){\bf background}
    \put(365.5,278.0){\boldmath $\btolshort$\unboldmath}
    \put(365.5,252.0){\boldmath $\btoccbartolspshort$\unboldmath}
    \put(365.5,226.0){\boldmath $\ctolspshort$\unboldmath}
    \put(199.0,366.0){\Large\boldmath$y$\unboldmath}
    \put(199.0,186.0){\Large\boldmath$y$\unboldmath}
    \put(457.0,366.0){\Large\boldmath$y$\unboldmath}
    \put(-4.5,463.0){\rotninety\rotninety\rotninety\Large\bf Entries}
    \put(-8.0,285.0){\rotninety\rotninety\rotninety\Large\bf Entries}
    \put(246.5,463.0){\rotninety\rotninety\rotninety\Large\bf Entries}
  \end{picture}
  \vspace{-188pt}
  \caption{\label{incly.fig}
The $y$ distributions of (a) electron, (b) muon, and (c) slow pion 
tagged events at a centre-of-mass energy of 189 GeV 
are given by the points with error bars.  Also plotted 
is the expectation from the simulation together with the
contributions from background (open histogram)
and $\btol$ (cross-hatch), $\btoccbartolsp$ (wide diagonal 
hatch), and $\ctolsp$ decays (narrow diagonal hatch). Note that in the fit,
the tagged events are divided into subsamples of different bottom and
charm purities to allow an extraction of both the bottom and charm
asymmetries.
}
  \end{center}
\end{figure}

The forward-backward asymmetries for bottom and charm, 
$\AFBb$ and $\AFBc$,
are extracted from the data using 
an unbinned maximum likelihood fit to the charge signed
polar angle distribution of 
the thrust axis $\vec{T}$ in lepton and slow pion tagged events.
It is assumed that $\vec{T}$ is the axis
along which the primary quark-antiquark pair is emitted.
The quantity $y=-q \cos\theta_{T}$ is computed event by event, where $q$
is the charge of the lepton or slow pion, and the thrust direction
$\vec{T}$ is defined such that $\vec{T} \! \cdot \! \vec{p} > 0$,
with $\vec{p}$ being the 
momentum of the jet containing the lepton or slow pion.
The inclusive $y$ distributions 
for electron, muon, and slow pion tagged events 
at $189\ \GeV$ are presented in Figure~\ref{incly.fig}.

In the fit, both
the bottom and charm asymmetries are to be determined.
Therefore, the fit uses for each event its probabilities 
to be a correctly tagged $\bb$ or $\cc$ event or background, 
as determined from the simulation:  The events are divided into 
several subsamples according to their $\netb$, $\netbc$, $\netc$, and $|y|$ 
values.  Three bins are used for each quantity, making a total of $81$ 
subsamples separately for both electron and muon tagged events and $27$ 
subsamples for the slow pion tagged events. These subsamples have
different bottom and charm purities and are fitted simultaneously.

The cross-section for producing a $\qq$ pair is assumed to 
depend on $y$ according to 
\begin{equation}
\label{assumedxs.eqn}
    \frac{{\rm d}\sigma_{\qq}}{{\rm d}y} 
  \sim 
    1 + y^2 + \frac{8}{3}\,\AFB^{\Pq}\, y \ .
\end{equation}
The likelihood ${\cal L}_{\rm sub}$ for one subsample is given by
\begin{equation}
   \ln {\cal L}_{\rm sub}
 = \sum_{\rm candidates} \! \ln \left(1 + y^2 + \frac{8}{3} \,
   \AFBobs \, y \right) \ ,
\end{equation}
where the sum is taken over all candidates in the subsample. The total 
likelihood is then given by the product of the likelihoods of all 
subsamples. The expected observed asymmetry $\AFBobs$ in each subsample is 
computed as
\begin{equation} 
\label{afbobs.eqn}
\AFB^{obs}(\netb, \netbc, \netc, |y|) = 
        \sum_{i=1}^{5} f_i(\netb, \netbc, \netc, |y|) \, \AFB^i \ .
\end{equation}
In this equation $f_i$ denotes the predicted fraction of leptons or slow pions
from source $i$, and $\AFB^i$ is the corresponding asymmetry: 
\begin{equation}
 \left\{
  \begin{array}{c@{\ =\ }r@{\,}cl}
 \AFB^1 & ( 1-2 \mix^{\rm eff}_1)     & \AFBb & \qquad {\rm for\ } \btol\,, \\
 \AFB^2 & -\, ( 1-2 \mix^{\rm eff}_2) & \AFBb & \qquad {\rm for\ } \btoctolsp\,, \\
 \AFB^3 & ( 1-2 \mix^{\rm eff}_3)     & \AFBb & \qquad {\rm for\ } \btocbartolsp\,, \\
 \AFB^4 &  -              & \AFBc & \qquad {\rm for\ } \ctolsp\,,\ {\rm and} \\
 \AFB^5 &                   &    0    & \qquad {\rm for\ background,} \\
\end{array} \right.
\label{equ:asyval}
\end{equation}
where $\mix_i^{\rm eff}$ is the effective $\PB\!-\!\PaB$ mixing parameter.
The fractions $f_i$ have been calculated from the Monte Carlo simulation  
and depend on the mis-identification probability in each 
bin of $\netb$, $\netbc$, $\netc$, and $|y|$, 
on the production rates of bottom and charm quarks,
on the semileptonic branching ratios of heavy hadrons, and on the hadronisation
fractions $f(\Pb\to[\Pc,\overline{\Pc}]\to\PDstpm)$ and $f(\Pc\to\PDstp)$.
Variations in the sample composition with $|\cos\theta_{T}|$ are
taken into account since the $f_i$ are binned in $|y|$.
As described in Section~\ref{sec:afb}, this 
likelihood fit has the advantage that the $|\cos\theta_{T}|$ dependence 
of the efficiency for identifying leptons and slow pions
is not needed explicitly.
The effective mixing parameters $\mix^{\rm eff}_1$, $\mix^{\rm eff}_2$, and
$\mix^{\rm eff}_3$ are determined from the simulation for each selected
subsample of events.  
The mixing parameter $\mix=0.118\pm0.006$~\cite{bib-PDG1998} 
used in the simulation for inclusive $\bb$ events is taken as an 
external input. The small non-zero contributions 
to the observed asymmetry from prompt leptons and slow pions from 
radiative and four-fermion events are accounted for, but left out of the 
above list for simplicity.
The assumption that the backgrounds from mis-identified leptons and slow pions
do not contribute to 
the observed asymmetry has been checked and will be discussed in 
Section~\ref{systunc_physics.subsubsec}.

The fit is done separately for the data taken at $133\ \GeV$, 
161 $\GeV$, 172 $\GeV$, 183 $\GeV$, and 189~$\GeV$. 
In Table~\ref{tab:lepton_results}, the
numbers of lepton and slow pion tagged events are given, and the results of
the fit for the bottom and charm asymmetries are summarised 
for each energy point together with the errors and correlations.
The systematic errors
have been evaluated as described in Section~\ref{sec:lept-sys}.

\begin{table}[htb]
\begin{center}
\begin{tabular}{|c||c|c|c||r@{}c@{}l|c||c|}
\hline
& & & Slow & \multicolumn{3}{c|}{Measured} & & \multicolumn{1}{c|}{Predicted} \\
Energy & $\!$Electrons$\!$ & $\!$Muons$\!$ & $\!$pions$\!$ &
\multicolumn{3}{c|}{asymmetries} & Correlation & 
\multicolumn{1}{c|}{asymmetries} \\
\hline\hline
133 \GeV 
         & \enspace 48 & \enspace 69 & \enspace 82
         & $\begin{array}{@{}r@{\,=\,}l@{\,}}
             \AFBb & -0.01 \rule{0pt}{3ex} \\
             \AFBc & \phantom{-} 0.50 \rule[-1.5ex]{0pt}{4.5ex} 
            \end{array}$
         & $\begin{array}{@{}c@{}}
             ^{+0.37}_{-0.35} \rule{0pt}{3ex} \\
             ^{+0.31}_{-0.32} \rule[-1.5ex]{0pt}{4.5ex} 
            \end{array}$
         & $\begin{array}{@{\,}l@{}}
             \pm\, 0.18 \rule{0pt}{3ex} \\
             \pm\, 0.13 \rule[-1.5ex]{0pt}{4.5ex} 
            \end{array}$
         & +14\% 
         & $\!\begin{array}{@{}r@{\,=\,}l@{}}
             \AFBbSM & 0.48 \rule{0pt}{3ex} \\
             \AFBcSM & 0.69 \rule[-1.5ex]{0pt}{4.5ex} 
            \end{array}\!$ \\
\hline
161 \GeV 
         & \enspace 16 & \enspace 37 & \enspace 42
         & $\begin{array}{@{}r@{\,=\,}l@{\,}}
             \AFBb & \phantom{-} 0.18 \rule{0pt}{3ex} \\
             \AFBc & \phantom{-} 0.83 \rule[-1.5ex]{0pt}{4.5ex} 
            \end{array}$
         & $\begin{array}{@{}c@{}}
             ^{+0.56}_{-0.52} \rule{0pt}{3ex} \\
             ^{+0.59}_{-0.60} \rule[-1.5ex]{0pt}{4.5ex} 
            \end{array}$
         & $\begin{array}{@{\,}l@{}}
             \pm\, 0.15 \rule{0pt}{3ex} \\
             \pm\, 0.12 \rule[-1.5ex]{0pt}{4.5ex} 
            \end{array}$
         & +14\% 
         & $\!\begin{array}{@{}r@{\,=\,}l@{}}
             \AFBbSM & 0.55 \rule{0pt}{3ex} \\
             \AFBcSM & 0.69 \rule[-1.5ex]{0pt}{4.5ex} 
            \end{array}\!$ \\
\hline
172 \GeV 
         & \enspace 14 & \enspace 29 & \enspace 20
         & $\begin{array}{@{}r@{\,=\,}l@{\,}}
             \AFBb & \phantom{-} 0.55 \rule{0pt}{3ex} \\
             \AFBc & \phantom{-} 0.67 \rule[-1.5ex]{0pt}{4.5ex} 
            \end{array}$
         & $\begin{array}{@{}c@{}}
             ^{+0.85}_{-0.87} \rule{0pt}{3ex} \\
             ^{+0.49}_{-0.54} \rule[-1.5ex]{0pt}{4.5ex} 
            \end{array}$
         & $\begin{array}{@{\,}l@{}}
             \pm\, 0.18 \rule{0pt}{3ex} \\
             \pm\, 0.12 \rule[-1.5ex]{0pt}{4.5ex} 
            \end{array}$
         & +23\% 
         & $\!\begin{array}{@{}r@{\,=\,}l@{}}
             \AFBbSM & 0.56 \rule{0pt}{3ex} \\
             \AFBcSM & 0.67 \rule[-1.5ex]{0pt}{4.5ex} 
            \end{array}\!$ \\
\hline
183 \GeV 
         & \enspace 77 & 104 & 126
         & $\begin{array}{@{}r@{\,=\,}l@{\,}}
             \AFBb & \phantom{-} 0.59 \rule{0pt}{3ex} \\
             \AFBc & \phantom{-} 0.56 \rule[-1.5ex]{0pt}{4.5ex} 
            \end{array}$
         & $\begin{array}{@{}c@{}}
             ^{+0.29}_{-0.31} \rule{0pt}{3ex} \\
             ^{+0.27}_{-0.28} \rule[-1.5ex]{0pt}{4.5ex} 
            \end{array}$
         & $\begin{array}{@{\,}l@{}}
             \pm\, 0.12 \rule{0pt}{3ex} \\
             \pm\, 0.11 \rule[-1.5ex]{0pt}{4.5ex} 
            \end{array}$
         & +23\% 
         & $\!\begin{array}{@{}r@{\,=\,}l@{}}
             \AFBbSM & 0.57 \rule{0pt}{3ex} \\
             \AFBcSM & 0.67 \rule[-1.5ex]{0pt}{4.5ex} 
            \end{array}\!$ \\
\hline
189 \GeV 
         & 190 & 269 & 359
         & $\begin{array}{@{}r@{\,=\,}l@{\,}}
             \AFBb & \phantom{-} 0.28 \rule{0pt}{3ex} \\
             \AFBc & \phantom{-} 0.52 \rule[-1.5ex]{0pt}{4.5ex} 
            \end{array}$
         & $\begin{array}{@{}c@{}}
             \pm 0.21 \rule{0pt}{3ex} \\
             ^{+0.18}_{-0.19} \rule[-1.5ex]{0pt}{4.5ex} 
            \end{array}$
         & $\begin{array}{@{\,}l@{}}
             \pm\, 0.12 \rule{0pt}{3ex} \\
             \pm\, 0.11 \rule[-1.5ex]{0pt}{4.5ex} 
            \end{array}$
         & +19\% 
         & $\!\begin{array}{@{}r@{\,=\,}l@{}}
             \AFBbSM & 0.58 \rule{0pt}{3ex} \\
             \AFBcSM & 0.66 \rule[-1.5ex]{0pt}{4.5ex} 
            \end{array}\!$ \\
\hline
\end{tabular}
\caption{\label{tab:lepton_results} 
For each centre-of-mass energy, the numbers of lepton and 
slow pion tagged events, the results for the bottom and charm asymmetries as 
measured with the lepton and slow pion tag, and their correlation are listed. 
The first error is statistical, and the second systematic.
In the last column, the Standard Model predictions for the asymmetries
are given.
}
\end{center}
\end{table}

As a cross-check, the fit is also performed on the calibration data taken at
the $\PZz$ peak in the years 1996, 1997, and 1998.  The results of this
fit are consistent with the average of LEP1 and SLD measurements given 
in~\cite{bib:ew}.

\subsection{Combination of the \boldmath$\AFBb$\unboldmath{ }Measurements}
\label{combination.subsec}

By construction, the samples used in the hemisphere charge analysis
do not have any events in common with the lepton or slow pion tagged samples.
The bottom asymmetry as measured with the vertex tag depends
on the assumed value of the charm asymmetry, while the fit to the lepton 
and slow pion tagged events yields results for the bottom and charm asymmetries
with a non-zero correlation.  This dependence on the charm asymmetry has to be
taken into account when combining the two bottom asymmetry measurements.
The vertex tag measurement is applied as a
constraint in the fit to the lepton and slow pion 
tagged events that is described in 
Section~\ref{fit.subsec}. Since the bottom and charm asymmetries are
correlated in the fit to the lepton and slow pion 
tagged events, both the fitted bottom
and charm asymmetries are expected to change in the constrained fit; but the
error on the charm asymmetry is essentially unchanged.

\begin{figure}[p]
\vspace{45pt}
\begin{center}
\end{center}
\vspace{-110pt}
\begin{center}
\epsfxsize=0.6\textwidth
\epsfbox[0 0 594 400]{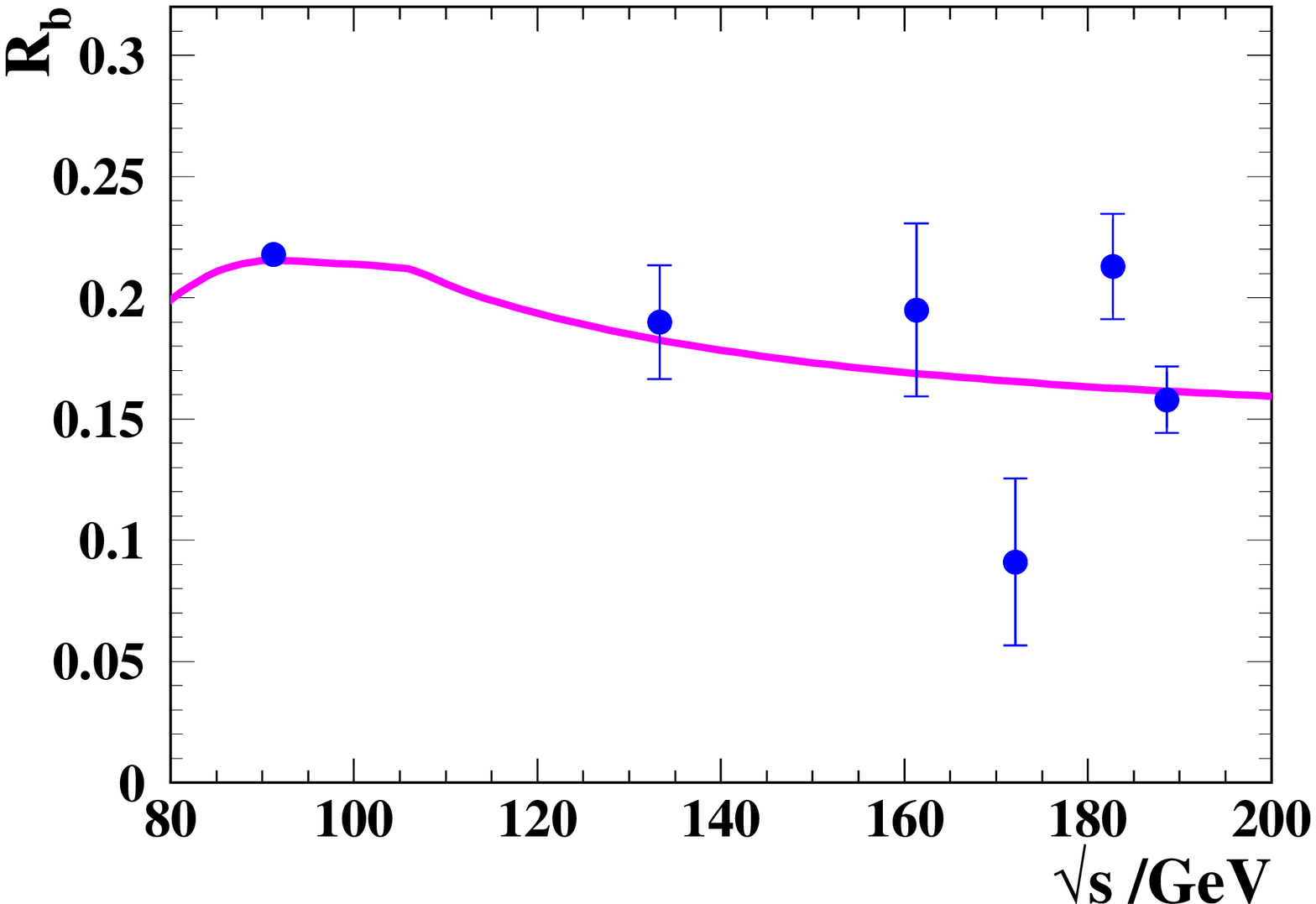}
\put(-238,160){\bf (a)}
\put(-161,50){\Large\bf OPAL}
\end{center}
\begin{center}
\epsfxsize=0.6\textwidth
\epsfbox[0 0 594 400]{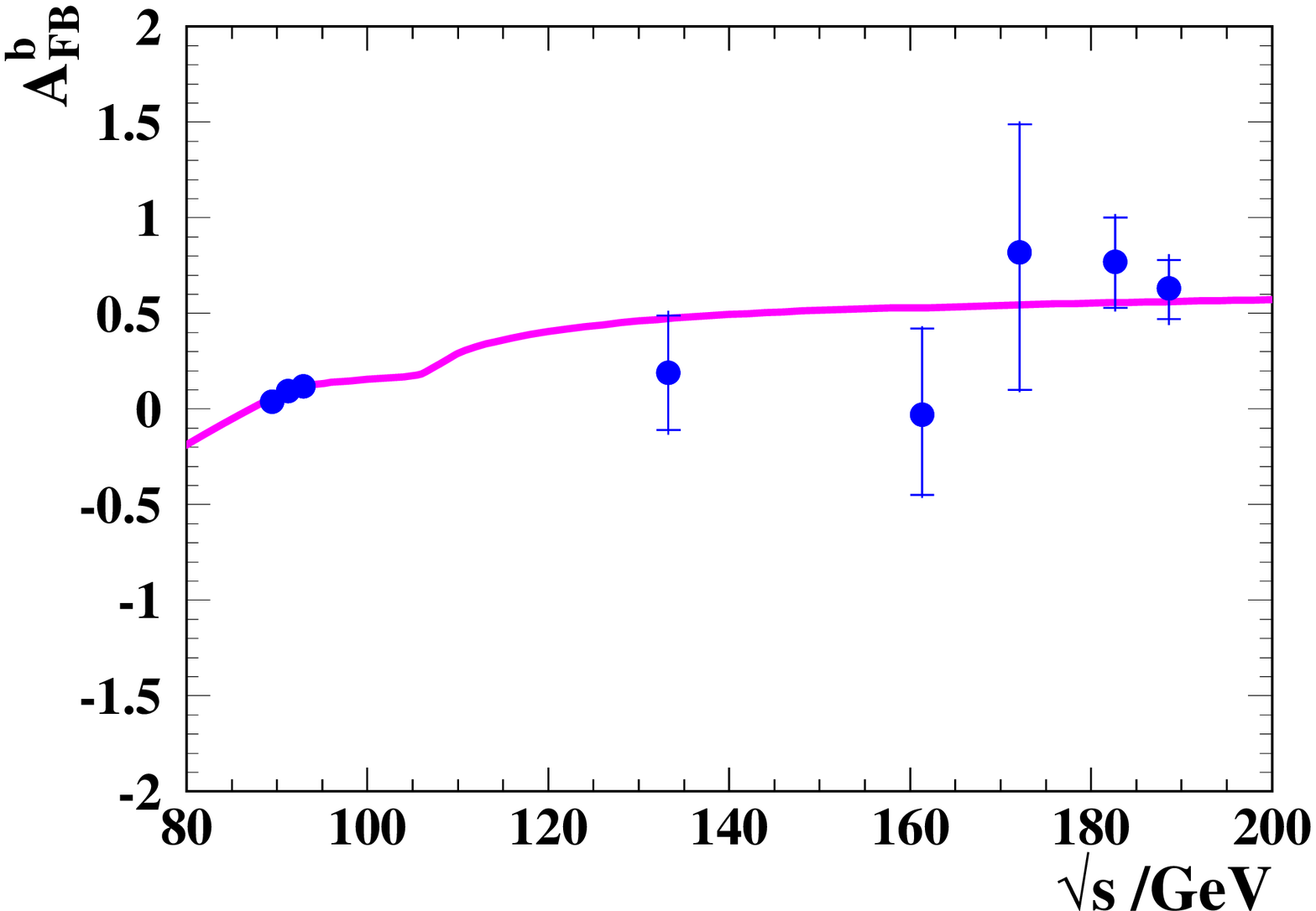}
\put(-238,160){\bf (b)}
\put(-161,50){\Large\bf OPAL}
\end{center}
\begin{center}
\epsfxsize=0.6\textwidth
\epsfbox[0 0 594 400]{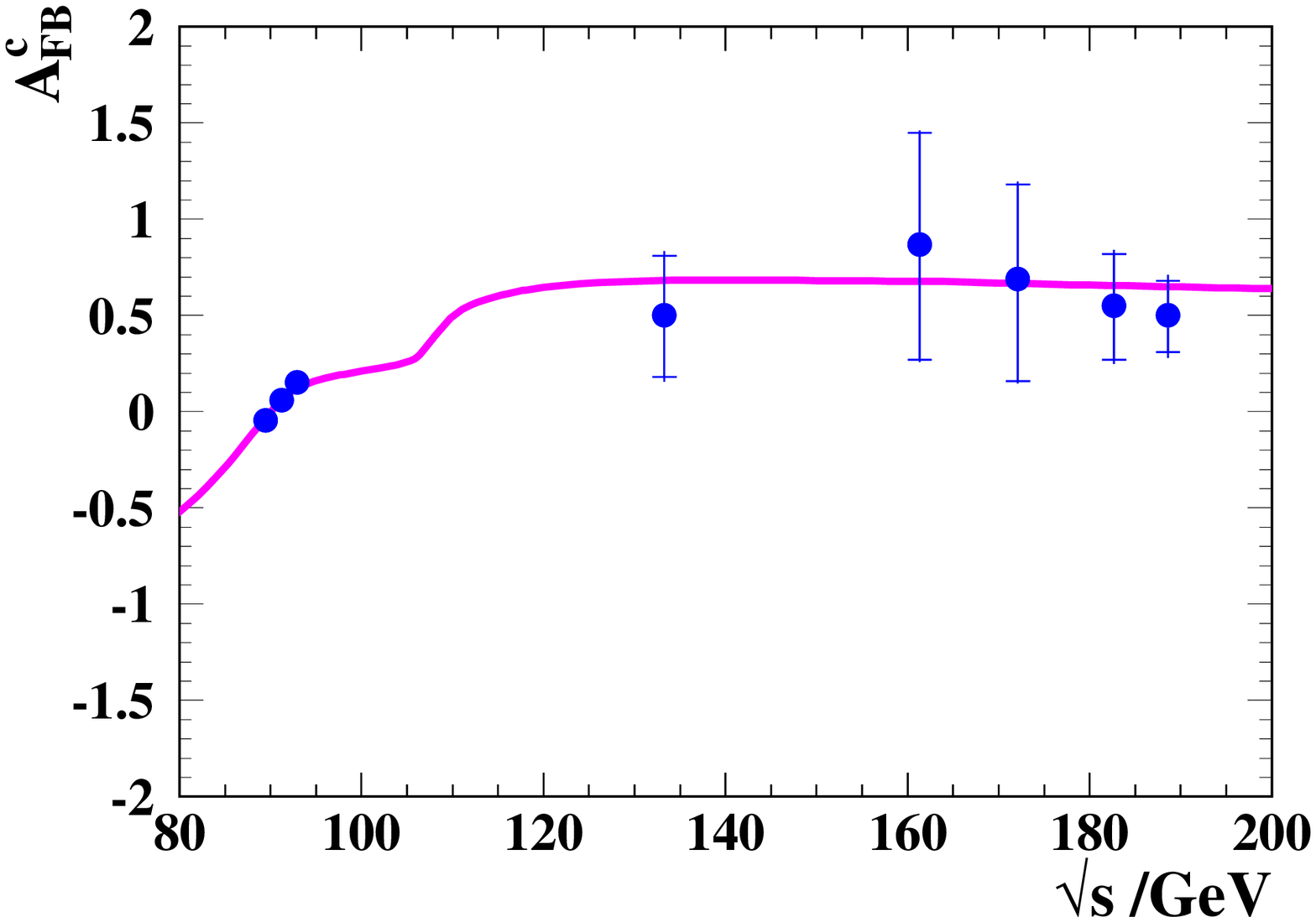}
\put(-238,160){\bf (c)}
\put(-161,50){\Large\bf OPAL}
\end{center}
\caption{The $\Rb$, $\AFBb$, and $\AFBc$ measurements are compared to the
Standard Model predictions. The statistical and the total errors of the 
measurements are indicated.  The measurements at $\sqrt{s}\approx m_{\PZz}$
have been taken from~\cite{bib-Rb,bib-OPALlafb,bib-AFBjetc,bib-AFBD}; 
here, the measurement errors are smaller than the points.
The behaviour of the curve at $\sqrt{s}\approx 110$ GeV is due to the
cut $\sqrt{s'/s} > 0.85$.}
\label{fig:results}
\end{figure}

The results for $\AFBb$ and $\AFBc$ are listed in 
Table~\ref{combinedresults.table} and are also shown in 
Figure~\ref{fig:results} together with the Standard Model expectations
as a function of the centre-of-mass energy. Good agreement is observed
between the measurements and the predictions from ZFITTER.

\begin{table}[htb]
\begin{center}
\begin{tabular}{|c||r@{}c@{}l|c||c|}
\hline
& \multicolumn{3}{c|}{Measured} & & \multicolumn{1}{c|}{Predicted} \\
Energy & \multicolumn{3}{c|}{asymmetries} & Correlation & 
\multicolumn{1}{c|}{asymmetries} \\
\hline\hline
133 \GeV 
         & $\begin{array}{@{}r@{\,=\,}l@{\,}}
             \AFBb & \phantom{-} 0.19 \rule{0pt}{3ex} \\
             \AFBc & \phantom{-} 0.50 \rule[-1.5ex]{0pt}{4.5ex} 
            \end{array}$
         & $\begin{array}{@{}c@{}}
             \pm 0.30 \rule{0pt}{3ex} \\
             ^{+0.31}_{-0.32} \rule[-1.5ex]{0pt}{4.5ex} 
            \end{array}$
         & $\begin{array}{@{\,}l@{}}
             \pm\, 0.12 \rule{0pt}{3ex} \\
             \pm\, 0.13 \rule[-1.5ex]{0pt}{4.5ex} 
            \end{array}$
         & +20\% 
         & $\!\begin{array}{@{}r@{\,=\,}l@{}}
             \AFBbSM & 0.48 \rule{0pt}{3ex} \\
             \AFBcSM & 0.69 \rule[-1.5ex]{0pt}{4.5ex} 
            \end{array}\!$ \\
\hline
161 \GeV 
         & $\begin{array}{@{}r@{\,=\,}l@{\,}}
             \AFBb & -0.03 \rule{0pt}{3ex} \\
             \AFBc & \phantom{-} 0.87 \rule[-1.5ex]{0pt}{4.5ex} 
            \end{array}$
         & $\begin{array}{@{}c@{}}
             ^{+0.45}_{-0.42} \rule{0pt}{3ex} \\
             ^{+0.58}_{-0.60} \rule[-1.5ex]{0pt}{4.5ex} 
            \end{array}$
         & $\begin{array}{@{\,}l@{}}
             \pm\, 0.11 \rule{0pt}{3ex} \\
             \pm\, 0.12 \rule[-1.5ex]{0pt}{4.5ex} 
            \end{array}$
         & +23\% 
         & $\!\begin{array}{@{}r@{\,=\,}l@{}}
             \AFBbSM & 0.55 \rule{0pt}{3ex} \\
             \AFBcSM & 0.69 \rule[-1.5ex]{0pt}{4.5ex} 
            \end{array}\!$ \\
\hline
172 \GeV 
         & $\begin{array}{@{}r@{\,=\,}l@{\,}}
             \AFBb & \phantom{-} 0.82 \rule{0pt}{3ex} \\
             \AFBc & \phantom{-} 0.69 \rule[-1.5ex]{0pt}{4.5ex} 
            \end{array}$
         & $\begin{array}{@{}c@{}}
             ^{+0.67}_{-0.72} \rule{0pt}{3ex} \\
             ^{+0.49}_{-0.53} \rule[-1.5ex]{0pt}{4.5ex} 
            \end{array}$
         & $\begin{array}{@{\,}l@{}}
             \pm\, 0.14 \rule{0pt}{3ex} \\
             \pm\, 0.12 \rule[-1.5ex]{0pt}{4.5ex} 
            \end{array}$
         & +27\% 
         & $\!\begin{array}{@{}r@{\,=\,}l@{}}
             \AFBbSM & 0.56 \rule{0pt}{3ex} \\
             \AFBcSM & 0.67 \rule[-1.5ex]{0pt}{4.5ex} 
            \end{array}\!$ \\
\hline
183 \GeV 
         & $\begin{array}{@{}r@{\,=\,}l@{\,}}
             \AFBb & \phantom{-} 0.77 \rule{0pt}{3ex} \\
             \AFBc & \phantom{-} 0.55 \rule[-1.5ex]{0pt}{4.5ex} 
            \end{array}$
         & $\begin{array}{@{}c@{}}
             ^{+0.23}_{-0.24} \rule{0pt}{3ex} \\
             ^{+0.27}_{-0.28} \rule[-1.5ex]{0pt}{4.5ex} 
            \end{array}$
         & $\begin{array}{@{\,}l@{}}
             \pm\, 0.10 \rule{0pt}{3ex} \\
             \pm\, 0.11 \rule[-1.5ex]{0pt}{4.5ex} 
            \end{array}$
         & +34\% 
         & $\!\begin{array}{@{}r@{\,=\,}l@{}}
             \AFBbSM & 0.57 \rule{0pt}{3ex} \\
             \AFBcSM & 0.67 \rule[-1.5ex]{0pt}{4.5ex} 
            \end{array}\!$ \\
\hline
189 \GeV 
         & $\begin{array}{@{}r@{\,=\,}l@{\,}}
             \AFBb & \phantom{-} 0.63 \rule{0pt}{3ex} \\
             \AFBc & \phantom{-} 0.50 \rule[-1.5ex]{0pt}{4.5ex} 
            \end{array}$
         & $\begin{array}{@{}c@{}}
             ^{+0.15}_{-0.16} \rule{0pt}{3ex} \\
             ^{+0.18}_{-0.19} \rule[-1.5ex]{0pt}{4.5ex} 
            \end{array}$
         & $\begin{array}{@{\,}l@{}}
             \pm\, 0.10 \rule{0pt}{3ex} \\
             \pm\, 0.11 \rule[-1.5ex]{0pt}{4.5ex} 
            \end{array}$
         & +29\% 
         & $\!\begin{array}{@{}r@{\,=\,}l@{}}
             \AFBbSM & 0.58 \rule{0pt}{3ex} \\
             \AFBcSM & 0.66 \rule[-1.5ex]{0pt}{4.5ex} 
            \end{array}\!$ \\
\hline
\end{tabular}
\caption{\label{combinedresults.table}
The results of the combined $\AFBb$ and $\AFBc$ measurements together with the 
Standard Model predictions from ZFITTER.  The first error is statistical, 
and the second systematic.}
\end{center}
\end{table}

In all asymmetry measurements, $\Rc$ and $\Rb$ are 
fixed to the Standard Model values as predicted by 
ZFITTER. The dependence of the measured values on the assumed values of
$\Rb$ and $\Rc$ is parametrised as 
\begin{equation}
  \begin{array}{r@{\;}l}
    \Delta \AFBb =   a^{\Pb}(\Rb) \; 
                     \frac{\displaystyle\Delta \Rb}{\displaystyle
                     R_{\rm b}^{\rm SM}}
                   + a^{\Pb}(\Rc) \; 
                     \frac{\displaystyle\Delta \Rc}{\displaystyle
                     R_{\rm c}^{\rm SM}}
    & \;  \rule{0pt}{4ex}\\
    \Delta \AFBc =   a^{\Pc}(\Rb) \; 
                     \frac{\displaystyle\Delta \Rb}{\displaystyle
                     R_{\rm b}^{\rm SM}}
                   + a^{\Pc}(\Rc) \; 
                     \frac{\displaystyle\Delta \Rc}{\displaystyle
                     R_{\rm c}^{\rm SM}}
    & ,\rule{0pt}{4ex}
  \end{array} 
\end{equation}
where $\Delta \Rq = \Rq-R_{\rm q}^{\rm SM}$ for q=b,\,c.
The Standard Model values for $\Rb$ and $\Rc$ are given in 
Table~\ref{Rqcoeff.table} together with the values of the coefficients
$a^{\rm q}(R_{\Pq'})$.
\begin{table}[htb]
\begin{center}
$\begin{array}{|c||c|c||c|c|c|c|}
\hline
{\rm energy} & R_{\rm b}^{\rm SM} & R_{\rm c}^{\rm SM} & 
a^{\rm b}(\protect{\Rb}) & a^{\rm b}(\protect{\Rc}) &
a^{\rm c}(\protect{\Rb}) & a^{\rm c}(\protect{\Rc}) \rule[-1.5ex]{0pt}{4.5ex}\\
\hline\hline
133\ \GeV
         & 0.184 & 0.223
         & -0.24 & +0.12 & +0.06 & -0.28  \rule[-1.5ex]{0pt}{4.5ex}\\
\hline
161\ \GeV
         & 0.171 & 0.244
         & -0.07 & -0.06 & +0.14 & -0.86  \rule[-1.5ex]{0pt}{4.5ex}\\
\hline
172\ \GeV
         & 0.168 & 0.249
         & -0.64 & +2.48 & +0.07 & -0.11  \rule[-1.5ex]{0pt}{4.5ex}\\
\hline
183\ \GeV
         & 0.165 & 0.253
         & -0.40 & +0.21 & +0.07 & -0.36  \rule[-1.5ex]{0pt}{4.5ex}\\
\hline
189\ \GeV
         & 0.164 & 0.255
         & -0.38 & +0.23 & +0.10 & -0.30  \rule[-1.5ex]{0pt}{4.5ex}\\
\hline
\end{array}$ 
\caption{The Standard Model predictions for the 
partial hadronic decay widths $\Rb$ and $\Rc$ and the dependence of the
results on their values are given.
The coefficients apply to the 
combined measurements.  The symbols are defined in the text.}
\label{Rqcoeff.table} 
\end{center}
\end{table}

Any systematic error from a common source (see Section~\ref{systunc.sec} for
the description of systematic uncertainties) is treated as fully correlated
between the two measurements.  However, there are large contributions to the
systematic error that affect only one of the analyses, such that systematic
errors are clearly no limitation to the combination of the two results.

\section{Systematic Errors}
\label{systunc.sec}

In this section, the evaluation of the systematic 
errors for the analyses presented here is discussed.
Both the secondary vertex and the lepton and slow pion tags depend
crucially on the knowledge of the detector resolution for reconstructing
charged particle tracks.
Most of the other main systematic errors 
are independent for the two measurements.

\subsection{Systematic Errors on \boldmath$\Rb$ and $\AFBb$\unboldmath{ }with
the Secondary Vertex Tag}
\label{sec:sys}

In Table~\ref{tab-systv}, a breakdown of the systematic error is given
for both the $\Rb$ and $\AFBb$ measurements with the vertex tag
at 189 GeV centre-of-mass energy.
The systematic errors considered for these measurements are
described in the following paragraphs.

\begin{table}[htb]
\begin{center}
\begin{tabular}{|l||r|r|}
 \hline
    \multicolumn{1}{|c||}{Error source}
  & \multicolumn{1}{c|}{$\Delta R_{\rm b}$} 
  & \multicolumn{1}{c|}{$\Delta A^{\rm b}_{\rm FB}$} \\
\hline\hline
Event selection  & 0.0038 & 0.036 \\
Final state QCD corrections & \multicolumn{1}{c|}{$-$} & 0.011 \\
b fragmentation & 0.0003 & 0.004\\
b decay multiplicity &0.0027 & 0.018\\
b hadron composition &0.0017 & 0.011\\
b lifetime           &$<$0.0001 & $<$0.001\\
c fragmentation      &$<$0.0001 & 0.001\\
c decay multiplicity & $<$0.0001 & 0.003\\
c hadron composition & 0.0011 & 0.011\\
c lifetime           &$<$0.0001 & $<$0.001\\ 
Four-fermion background       & 0.0009 & $<$0.001   \\
Monte Carlo statistics        & 0.0015& 0.031 \\
Track reconstruction &0.0051 & 0.057\\ \hline\hline
Total systematic error  &0.0074 & 0.078 \\ \hline 
\end{tabular}
\caption{\label{tab-systv}
Systematic error breakdown 
for the $\Rb$ and the $\AFBb$ measurements with the vertex tag at $\sqrt{s}$ = 189 GeV. 
Similar uncertainties have been determined for the other centre-of-mass
energies.}
\end{center}
\end{table}


\subsubsection{Event Selection}
\label{evselsys.subsubsec}
The bias on the measurement of $\Rb$ from the selection of non-radiative
events has already been discussed in
Section~\ref{sec:rb}. A correction has been applied, and an uncertainty of
100$\%$ is assigned to the correction.
The uncertainty on the corrections for 
the interference between initial and final 
state radiation results in a systematic error of 0.5\%
on $\Rb$. 
The contamination from radiative events 
has the effect of decreasing
the measured asymmetry by $3\%$.  The value is corrected
accordingly, and a $3\%$
systematic error is assigned.
The likelihood fit is based on the assumption 
that the shape of the tagging efficiency as a function of
$|\cos\theta_T|$ is the same for all flavours, which 
might not be true at the edges of 
the acceptance.  
The effect has been estimated by dividing the sample in bins of
$|\cos\theta_T|$, over which the above assumption is valid,
determing the asymmetry in each bin independently and comparing their
average with the reference result.
It has been found to affect the asymmetry by less
than 1$\%$, and is thus neglected.

\subsubsection{Final State QCD Corrections}
\label{qcdcorr.subsubsec}
Final state QCD corrections are included in the calculation of the 
predictions with which the asymmetry measurements are compared.
However, the experimental
event selection is less efficient for events with hard gluon radiation
due to the cuts on the decay length significance (for the vertex tag)
and the lepton or slow pion momentum (for the lepton and slow pion analysis).
The fraction of the QCD correction that has to be applied to the measurements
has been 
determined previously at LEP1 and is typically between $0.3$ and
$0.6$~\cite{bib-QCDcorrections}. The overall QCD corrections
at $\sqrt{s}=189\,\GeV$ have been determined from ZFITTER to be
0.015 for $\AFBb$ and 0.022 for $\AFBc$, respectively.
Half of this correction is assigned as a systematic error.

\subsubsection{Bottom and Charm Physics Modelling}
\label{bcphmod.subsubsec} 
Uncertainties in bottom and charm fragmentation and decay properties 
have been treated as follows:

\begin{list}{$\bullet$}{\setlength{\itemsep}{0ex}
                        \setlength{\parsep}{1ex}
                        \setlength{\topsep}{0ex}}
\item{\bf b fragmentation:} Although the mean scaled energy 
$2 \langle E_{\rm b} \rangle/\sqrt{s}$
of weakly decaying b hadrons is expected to change from LEP1 to LEP2 energies,
the Peterson fragmentation parameter $\epsilon^{\rm b}_P$, 
which describes one step in the fragmentation process, is assumed not 
to vary with energy.
Simulated $\bb$ events are reweighted within the range of 
$0.0030 < \epsilon^{\rm b}_P < 0.0048$~\cite{bib:lep_heavy},
which corresponds to a variation of the
mean scaled energy $2 \langle E_{\rm b} \rangle/\sqrt{s}$
of weakly decaying b hadrons in $\PZz$ decays in the range of
$2 \langle E_{\rm b} \rangle/\sqrt{s} = 0.702 \pm 0.008$~\cite{bib:lep_heavy}.


\item {\bf b decay multiplicity:} The charged decay multiplicity of 
hadrons containing a b quark is 
varied in the Monte Carlo simulation according 
to~\cite{bib:lep_heavy}.


\item {\bf b hadron composition:}
The tagging efficiency differs for the various b~hadron species.
The fractions of b~hadrons and their errors 
have been taken from~\cite{bib-PDG1998}.
The fractions $f({\rm B^0 + B^+})$ and  $f({\rm B^0_s})$ are varied
independently within their errors, and their variation 
is compensated by the b baryon fraction.


\item {\bf b hadron lifetimes:} The lifetimes of the
different b hadrons are varied in the Monte Carlo by their
errors according to~\cite{bib-PDG1998}.


\item  {\bf c fragmentation:}  
$\rm c\overline{c}$ Monte Carlo events 
are reweighted by varying the
Peterson fragmentation parameter $\epsilon^{\rm c}_P$ in the range of
$0.025 < \epsilon^{\rm c}_P < 0.031$ ~\cite{bib:lep_heavy}. 
This corresponds to a variation of the
mean scaled energy $2 \langle E_{\rm c} \rangle/\sqrt{s}$ of weakly decaying 
c hadrons in $\PZz$ decays in the range of 
$2 \langle E_{\rm c} \rangle /\sqrt{s}= 0.484 \pm 0.008$.
 
\item  {\bf c decay multiplicity:} The average charged track multiplicities
of $\rm D^+$, $\rm D^0$ and $\rm D^+_s$ decays are varied in the Monte Carlo 
within the ranges of the experimental measurements~\cite{bib:mark3}.

\item  {\bf c hadron composition:}
The $\rm D^0$
fraction is written as 
$f(\PDz) = 1- f(\PDp) - f({\PDsp}) - f({\rm c}_{baryon}) $.
The last three parameters are varied independently by their errors
according to~\cite{bib:lep_heavy}
to evaluate the uncertainty on the charm efficiency. 

\item   {\bf c hadron lifetimes:} Charmed hadron 
lifetimes are varied within their experimental errors according to~\cite{bib-PDG1998}.

\end{list}

\subsubsection{\bf Four-fermion Background}
Above the WW production threshold,
the four-fermion background is largely dominated by W pairs.
The uncertainty in the W pair production cross-section is 
taken into account and
has been found to have a negligible systematic effect.
The background from W pairs 
has the highest probability to be accepted in the tagged sample 
when one or both W bosons 
decay into a final state containing a charm quark. 
The systematic error on the W pair tagging efficiency is estimated by 
varying the charm physics modelling parameters as described above.
The effect of the detector resolution is also taken into account, 
as described in Section~\ref{detres.subsubsec}.

\subsubsection{Monte Carlo Statistics}
Tagging efficiencies and charge 
identification probabilities are varied by the statistical error arising
from the finite number of Monte Carlo simulated events.

\subsubsection{Track Reconstruction}
\label{detres.subsubsec}

The effect of the detector resolution on the track 
parameters is estimated by degrading or improving the resolution of 
all tracks in the Monte Carlo simulation.
This is done by applying a single multiplicative scale factor to the 
difference between the reconstructed and true track parameters.
A $\pm 10 \%$ variation is applied independently
to the $r\phi$ and $rz$ track parameters.

In addition, the matching efficiency for assigning measurement points
in the silicon microvertex detector to the tracks is varied by $1\%$
in $r\phi$ and $3\%$ in $rz$.
The systematic errors resulting from the 
individual variations are summed in quadrature.

\subsection{Systematic Errors on \boldmath$\AFBb$ and 
$\AFBc$\unboldmath{ }with the Lepton and Slow Pion Tag}
\label{sec:lept-sys}
Three different groups of systematic errors have been considered:
those from detector effects, those related to physics models 
and external inputs used in the analysis, and effects 
introduced by the limited Monte Carlo statistics and the fitting procedure.
A list of all systematic errors is given in
Table~\ref{tab-syst}.
\begin{table}[htb]
\begin{center}
\begin{tabular}{|l|
                |c|c|}
\hline
\multicolumn{1}{|c||}{Error source} 
                                    & $\Delta\AFB^{\Pb}$ & $\Delta\AFB^{\Pc}$ \\
\hline
\hline

Track reconstruction
                          & \enspace 0.088\enspace & \enspace 0.070\enspace \\

Lepton and slow pion identification
                          & 0.009 & 0.013 \\

Input modelling           
                          & 0.045 & 0.022 \\

\hline

Fragmentation modelling   
                          & 0.024 & 0.033 \\

Semileptonic decay models 
                          & 0.007 & 0.005 \\

Branching ratios          
                          & 0.008 & 0.006 \\

$\PB\!-\!\PaB$ mixing                  
                          & 0.003 & 0.001 \\

Background asymmetries
                          & 0.024 & 0.066 \\

Final state QCD corrections
                          & 0.008 & 0.011 \\

\hline

Monte Carlo statistics    
                          & 0.061 & 0.032 \\

Fitting procedure
                          & 0.009 & 0.003 \\

\hline
\hline
Total systematic error    
                          & 0.122 & 0.111 \\
\hline
\end{tabular}
\caption{A breakdown of the systematic errors is given for 
the asymmetry measurement with leptons and slow pions at 189 GeV.
Similar systematic errors have been obtained for the measurements at the
other centre-of-mass energies above the $\PZz$ peak.
}
\label{tab-syst} 
\end{center}
\end{table}

\subsubsection{Detector Systematics}
The lepton and slow pion identification relies on the 
proper modelling of the detector response in the 
Monte Carlo simulation. 

\begin{list}{$\bullet$}{\setlength{\itemsep}{0ex}
                        \setlength{\parsep}{1ex}
                        \setlength{\topsep}{0ex}}
\item{\bf Track Reconstruction:}
The influence of the simulated detector resolution 
and matching efficiency for hits in the microvertex detector
is estimated as described in Section~\ref{detres.subsubsec}.

\item{\bf Lepton Identification:}
The fractions of misidentified electrons and muons 
are taken from Monte Carlo simulation.
The modelling of the input variables to the artificial neural network 
for electron identification has been studied on $\PZz$ calibration data
in a manner similar to that described 
in~\cite{bib-Rb}.  Differences between the data and the modelling in the
Monte Carlo simulation have been determined using a pure sample of electrons 
from photon conversions and an inclusive sample of tracks depleted in 
conversion electrons.  The 
dependence of the efficiency and background contamination
of the selected electron sample on these differences has been studied, and
the resulting uncertainties for each input have been added in quadrature.
In addition, it has been shown that there is no large dependence of the 
resulting systematic error on the position of the $\netel$ cut.  The
systematic error has been evaluated separately for each year of 
data taking at energies above the $\PZz$ peak; 
the uncertainty on the efficiency is around $10\%$, and that on the 
background contamination around $20\%$.

The performance of the muon tagging network has been studied on 
$\gamma\gamma\to\mu^+\mu^-$ events recorded at LEP energies above the
$\PZz$ peak
and on an inclusive 
sample of tracks from the $\PZz$ calibration data that fail the ``best match'' 
preselection criterion. The muon and background rates have been compared
between data and Monte Carlo simulation in bins of $\netmu$, and the 
simulation has been reweighted to match the data distribution.  
The reweighting factors
in each bin have been varied independently by their statistical errors, and 
the resulting variations of the measured asymmetries added in quadrature.
For the inclusive muon sample selected with a cut at $\netmu>0.65$, the
reweighting factors for muon signal and background are
$0.95 \pm 0.05$ and $0.97 \pm 0.04$, respectively.

Conversion candidates are explicitly removed from the sample of events used 
in the fit. The $\netcv$ output distribution is 
well described in the simulation.
The conversion finding efficiency is tested 
in the data using an algorithm which does not need the $\netel$ outputs
for the two tracks as input. The accuracy of this test is $18\%$, and
the conversion rate is varied by this amount to calculate the systematic
error.
The rate of non-prompt muons from pion and kaon decays in flight has been
studied previously in~\cite{bib-muonsel}, and has been 
found to be modelled to within $9\%$.

The four above systematic errors are added in quadrature to yield
the value given in Table~\ref{tab-syst}.

\item{\bf Slow Pion Efficiency:}
Because of the large backgrounds in the slow pion
sample, it is crucial to check that the slow pion
reconstruction is modelled correctly in the Monte Carlo simulation. The
efficiency has been studied by reconstructing slow pions in the $\PZz$
calibration data using the same cuts as described in 
Section~\ref{spsel.subsec}, except for the upper momentum cut
which has been lowered to 
$p_{\pi_s}^{\rm max}(\sqrt{s} = m_{\PZz}) = 4.17\ \GeV$.
The slow pion content
in this sample has been measured using a fit to the $\ptsq$ spectrum.
Functions which are found to describe the shape of the background
distribution well in the simulation are fitted to the $\ptsq$ sideband
from $0.03$ to $0.10\ \GeV^2$.  The number of signal slow pion events is then
determined by extrapolating the background estimate to $\ptsq=0\ \GeV^2$.
In the $\PZz$ calibration data, this fit has a relative statistical precision
of $6\%$, but an additional systematic
error of $23\%$ is assigned to cover biases resulting from the extrapolation
procedure and the particular choice of fit function.

Combining this result with the OPAL measured yield of $\PDstp$ mesons in 
hadronic $\PZz$ decays of
$\bar{n}_{\PZz\to\PDstp{\rm X}}
=0.1854 \pm 0.0041 \pm 0.0059 \pm 0.0069$~\cite{bib-Ties},
the efficiency of the slow pion reconstruction in $\PZz$ 
decays is measured to be $(37.3 \pm 9.0)\%$, where the relative errors of
$6\%$ and $23\%$ have been added in quadrature. This is consistent with the
value in the simulation of $(31.9 \pm 1.5({\rm stat.}))\%$.  
The fractional error on the slow pion efficiency is then assigned 
to the slow pion efficiencies at energies above the $\PZz$ peak, which are 
taken from the simulation.
In order to determine the systematic error on the asymmetry measurements, 
the expected number of slow pions in each of the different subsamples is varied 
according to the error on the efficiency, and balanced by the expected
number of background events so that the total number of tagged events is kept 
constant.  

No additional systematic error is assigned for the modelling
of the shape of the slow pion signal $\ptsq$ distribution
since the signal shape in $\PZz$ decays
has been found to be well described in the simulation.

\item{\bf Modelling of Artificial Neural Network Inputs:}
Each of the input distributions used in the flavour separation networks
has been compared between 
data and Monte Carlo simulation. The simulated distributions are reweighted
for each input variable in turn
to agree with the corresponding data distributions, and the analysis is
repeated with the weighted events.
The observed differences from the original fit
result are added in quadrature to yield the systematic 
uncertainty due to the modelling of the input variables. 
\end{list}

\subsubsection{Physics Systematics}
\label{systunc_physics.subsubsec}

The fragmentation of charm and bottom 
quarks and the momentum spectra of leptons emitted in their 
semileptonic decay are described 
using phenomenological models tuned to experimental data. 
Systematic errors introduced 
by models of heavy hadron semileptonic decays, by the use of 
externally measured inputs, and from the assumed asymmetries of background 
candidates are assessed as follows:

\begin{list}{$\bullet$}{\setlength{\itemsep}{0ex}
                        \setlength{\parsep}{1ex}
                        \setlength{\topsep}{0ex}}
\item{\bf Lifetimes:}
The lifetimes of weakly decaying b- and c-flavoured hadrons are varied as
described in Section~\ref{bcphmod.subsubsec}.  The resulting changes
in measured asymmetries are negligible.

\item{\bf Fragmentation:}
Bottom and charm fragmentation uncertainties are estimated as described in
Section~\ref{bcphmod.subsubsec}. 

\item{\bf Semileptonic Decay Models:}
Systematic effects due to the modelling of 
semileptonic decays of heavy hadrons are studied 
following the recommendations in~\cite{bib:lep_heavy}.
The lepton momentum spectra in the Monte Carlo simulation are reweighted
to different theoretical models, with ranges of parameters 
chosen such that the experimental errors are covered. 

\item{\bf Branching Ratios and Hadronisation Fractions:}
The values for the semileptonic branching ratios 
$B(\btol)$, $B(\btoctol)$, \mbox{$B(\btocbartol)$}, and $B(\ctol)$
are taken from~\cite{bib:lep_heavy}, and they are varied 
within their errors.
Similarly, the values for the hadronisation fractions
$f(\Pb\to\Pc\to\PDstp)$ and $f(\Pc\to\PDstp)$ 
are taken from~\cite{bib-Ties} and varied within their errors.

In the absence of any measurements, the hadronisation fraction 
$f(\Pb\to\Pac\to\PDstm)$ is taken from the simulation and varied
by $100\%$ in order to assess the systematic error.

\item{\bf \boldmath$\PB\!-\!\PaB$\unboldmath{ }Mixing:}
The $\PB\!-\!\PaB$ mixing parameter is taken as an external input. 
The value of $\mix = 0.118 \pm 0.006$~\cite{bib-PDG1998} is used,
and the variation within its error 
yields the systematic uncertainty listed in Table~\ref{tab-syst}.

\item{\bf Background Asymmetries:}
Prompt leptons and slow pions 
from four-fermion background in the selected event sample 
can affect the measured asymmetries. 
They may lead to a rather large observed forward-backward 
asymmetry, but their contribution to the overall sample is small.
The asymmetry from these events is varied within $\pm 0.5$ to determine
the systematic error.  The error due to the uncertainties on the bottom
and charm asymmetries in radiative events is negligible.

The asymmetry in the $y$ distribution of mis-identified lepton and 
slow pion candidates is assumed to be zero for any event type.
For lepton candidates, this assumption is verified within an accuracy of $2\%$ 
on an inclusive sample of tracks that pass the lepton momentum cuts and
do not pass the cut on $\Psig$.
For slow pions, a sample of tracks that pass the cuts on the slow pion
momentum and transverse momentum, but do not pass the cut on $\Psig$, is
used. The asymmetry of slow pion background is found to be zero with
an error of $4\%$.
The asymmetries from lepton and slow pion backgrounds are 
then varied within these
limits to assess the associated systematic errors.

All the above errors are added in quadrature to yield the
uncertainty given in Table~\ref{tab-syst}.

\item{\bf Final State QCD Corrections:}
As described in Section~\ref{qcdcorr.subsubsec}, half the QCD
correction to $\AFBb$ and $\AFBc$
as computed using ZFITTER is assigned as systematic
error.
\end{list}

\subsubsection{Monte Carlo Statistics}
An error arises from the limited statistics in the Monte Carlo
simulation that is used to predict the fractions $f_i$ for each subsample (see
Section~\ref{fit.subsec}).  This uncertainty has been evaluated by varying
for each subsample in turn the contribution from each source by its statistical
error, redoing the fit, and adding all observed differences in quadrature.

\subsubsection{Fitting Procedure}
The fitting procedure has been studied in a 
large number of simulated experiments with the same statistics as in the actual
measurements.
It has been found to be essentially unbiased for the 
large data samples recorded
at 183 GeV and 189 GeV. For the datasets at 133 GeV, 161 GeV, and
172 GeV, however, the statistical error from the fit has
been found to be underestimated by
a factor of up to $1.16$ (bottom) or $1.05$ 
(charm asymmetry), and
biases of up to $0.12\, \sigma$ and $0.06\, \sigma$ have 
been found for the fitted bottom and charm asymmetry, respectively, 
where in each case $\sigma$ denotes the statistical error.
The statistical errors are adjusted, 
the measurements are corrected according to the observed biases, 
and the full bias is treated as an additional
systematic error.

\subsubsection{Cross-Checks}
Consistent results are observed when the fit is repeated with the numbers of 
bins in $\netb$, $\netbc$, $\netc$, and $|y|$ varied independently between 
2 and 4.
In addition, all selection cuts have been varied, with no significant 
deviations in the results.
Consistent results have been found when the slow pion tagged events are not 
used in the fit.  In the 189 GeV data sample, the statistical error of the 
charm asymmetry increases from $0.18$ to $0.21$ without the information from 
slow pion tagged events, while the error on the bottom asymmetry stays the 
same.

\section{Conclusions}
\label{conclusions.sec}

Using data collected at centre-of-mass energies between $130 \,\GeV$ and
$189\,\GeV$ with the OPAL detector at LEP, the 
relative $\rm{ e^+e^- \rightarrow b \overline{b}}$ production rate 
and the forward-backward asymmetries in 
$\rm{ e^+e^- \rightarrow b \overline{b}}$ and 
$\rm{ e^+e^- \rightarrow c \overline{c}}$ production have been 
measured to be
\[
  \begin{array}{|c||c|r@{\,}c@{\,}l|c|}
    \hline
     {\rm Energy} & 
     \protect{\Rb} &
     \multicolumn{3}{c|}{\protect{\AFBb}} & 
     \protect{\AFBc} \rule[-1.5ex]{0pt}{4.5ex} \\
    \hline
    \hline
     133\ \GeV &
        0.190 \pm 0.023 \pm0.007 & 
        \phantom{-}0.19 & \pm\, 0.30 & \pm\, 0.12 &
        0.50 \; ^{+0.31}_{-0.32} \pm 0.13 \rule{0pt}{3.0ex} \\
     161\ \GeV &
        0.195 \pm 0.035 \pm 0.007 &
        -0.03 & \hspace{2pt}^{+0.45}_{-0.42}\hspace{-2pt} & \pm 0.11 & 
        0.87 \; ^{+0.58}_{-0.60} \pm 0.12 \rule{0pt}{2.6ex} \\
     172\ \GeV &
        0.091 \pm 0.034 \pm 0.005 & 
        \phantom{-}0.82 & \hspace{2pt}^{+0.67}_{-0.72}\hspace{-2pt} & \pm 0.14 &
        0.69 \; ^{+0.49}_{-0.53} \pm 0.12 \rule{0pt}{2.6ex} \\
     183\ \GeV &
        0.213 \pm 0.020 \pm 0.009 &
        \phantom{-}0.77 & \hspace{2pt}^{+0.23}_{-0.24}\hspace{-2pt} & \pm 0.10 &
        0.55 \; ^{+0.27}_{-0.28} \pm 0.11 \rule{0pt}{2.6ex} \\
     189\ \GeV &
        0.158 \pm 0.012 \pm 0.007 & 
        \phantom{-}0.63 & \hspace{2pt}^{+0.15}_{-0.16}\hspace{-2pt} & \pm 0.10 &
        0.50 \; ^{+0.18}_{-0.19} \pm 0.11 \rule[-1.5ex]{0pt}{4.1ex} \\
     \hline
  \end{array}
\]
where in each case, the first error is statistical and the second systematic. 
These values are illustrated in Figure~\ref{fig:results} together with 
the dependences on the centre-of-mass energy as predicted in the 
Standard Model.  For all measurements, good agreement is observed with 
the Standard Model expectation.
The measurements of $\bb$ production presented in this paper supersede
the previously published values of references~\cite{bib:rb_172,bib:rb_183}.

\newpage

\end{document}